\documentclass[12pt,onecolumn]{IEEEtran}
\usepackage{bbm}
\usepackage{amsmath}
\usepackage{amsfonts}
\usepackage{mathrsfs}
\usepackage{amssymb}
\usepackage{cases}
\usepackage{stmaryrd}
\usepackage{mathrsfs}
\usepackage{bbm}
\usepackage[dvips]{graphicx}
\usepackage{algorithm}
\usepackage{algorithmic} %用到的宏包，要自己改下
\usepackage{multirow}
\newtheorem{lemma}{Lemma}
\newtheorem{Th}{Theorem}
\newtheorem{remark}{Remark}

\newtheorem{definition}{Definition}
\IEEEoverridecommandlockouts

\begin{document}
\baselineskip 4.2ex
\title{Physical Layer Security in Millimeter Wave Cellular Networks}

\author{\IEEEauthorblockN{Chao Wang and Hui-Ming Wang,~\IEEEmembership{Member, IEEE}}
\thanks{C. Wang and H.-M. Wang are with the School of Electronic and Information Engineering, and also with
the MOE Key Lab for Intelligent Networks and Network Security,
Xi'an Jiaotong University, Xi'an, 710049, Shaanxi, China. Email:
{\tt wangchaoxuzhou@stu.xjtu.edu.cn} and {\tt xjbswhm@gmail.com}.
The contact author is Hui-Ming Wang.
%This work was supported by the NSFC
%under Grants No. 61102081, the Science Foundation
%for Innovation Research Group of China under Grant No. 61221063, the
%Specialized Research Fund for the Doctoral Program of Higher Education of
%China under Grant 20110201120013, the New Century Excellent Talents Support Fund under Grant NCET-13-0458,
%the Industrial Research Fund of Shaanxi Province under Grant 2012GY2-28.
}
%and Jia Xiao \thanks{Jia Xiao is with State Key Laboratory of Networking and Switching Technology, Beijing University of Posts and Telecommunications (BUPT), 100876,
%Peking, China. Email: {\tt  tonyalex2010@163.com}}
}

\maketitle
%Although the global optimal solution of a D.C. program can be found by many global optimization algorithms, for striking a balance between computational complexity and optimality,

\begin{abstract}
Recent researches show that millimeter wave (mmWave) communications can offer orders of magnitude increases in the cellular capacity. However, the \emph{secrecy} performance of a mmWave cellular network has not been investigated so far.
Leveraging the new path-loss and blockage models for mmWave channels, which are significantly different from the conventional microwave channel, this paper comprehensively studies the network-wide physical layer security performance of the downlink transmission in a mmWave cellular network under a stochastic geometry framework.
We first study the secure connectivity probability and the average number of perfect communication links per unit area in a noise-limited mmWave network for both non-colluding and colluding eavesdroppers scenarios, respectively. Then, we evaluate the effect of the artificial noise (AN) on the secrecy performance, and derive the analysis result of  average number of perfect communication links per unit area in an interference-limited mmWave network. Numerical results are demonstrated to show the network-wide secrecy performance, and provide interesting insights into how the secrecy performance is influenced by various network parameters: antenna array pattern, base station (BS) intensity, and AN power allocation, etc.
\end{abstract}


% \fi
\IEEEpeerreviewmaketitle

\section{Introduction}
Within the next 20 years, wireless data traffic can be anticipated to skyrocket 10,000 folds, spurred by the popularity of various intelligent devices. Conventional communication means are difficult to meet such incredible increase in the wireless data traffic. Millimeter wave cellular communication has received an increasing attention due to the large available bandwidth at millimeter wave frequencies \cite{5GMillimeter}.  Recent field measurements have shown the huge advantages of mmWave networks, compared with the conventional microwave network in band below 6 GHz \cite{5GMillimeter,outdoormmWave,Millimeterpotentials}.
Due to the small wavelength, the mmWave cellular network is different from the conventional  microwave network in the following ways: large number of antennas, sensitivity to blockages, and variable propagation laws, etc \cite{CoverageMillimeter}.  Recently, based on the real-world measurements in \cite{Millimeterpotentials}, spatial statistical models of the mmWave channel have been built in \cite{MillimeterChannelModeling}, which reveal the different path loss characteristics of the line-of-sight (LOS) and non-line-of-sight (NLOS) links. Under the new channel model, the network-wide performance of a mmWave cellular has attracted increasing attentions, and  many works have investigated the SINR distribution, coverage, and average ergodic rate of the network under a stochastic geometry framework \cite{AnalysisBlockage}-\cite{MultiTierMillimiterWave}. They show that the mmWave network has a great potential to provide tremendous data traffic increase.

All the above works have focused on the \emph{rate/reliability} performance of the mmWave network, however, its \emph{secrecy} performance has not been investigated so far. Given the ubiquitousness of wireless connections, an enormous
amount of sensitive and confidential information, e.g. financial data, electronic cryptography, and private video, have been transmitted via wireless channels. Thus, providing a secure service is one of the top priorities in the design and implementation of mmWave networks \cite{Safeguarding5G}. In this paper, we investigate the physical layer security performance of mmWave networks by adopting the stochastic geometry framework.
%All of works above resort to the simulation-based approach which does not lead to an elegant system analysis.

\subsection{Background}
Physical layer security has been identified as a promising strategy that exploits randomness of wireless medium to protect the confidential information from wiretapping \cite{Wyner}, \cite{Mukherjee2014Principles}. Recently,  multiple-antenna technology becomes a powerful tool for enhancing the physical layer security in random networks \cite{PHYsignalProcessing}. With the degrees of freedom provided by multiple antennas, the transmitter can adjust its antenna steering orientation to exploit the maximum directivity gain while reducing the signal leakage to eavesdroppers \cite{Secureconnectivity}-\cite{Geraci2014Physical}, or radiate the artificial nosie (AN) for jamming  potential eavesdroppers \cite{Goel:TWC08}-\cite{Enhancing}.
The secure connectivity, secrecy rate and secrecy outage with multi-antenna transmissions in wireless random networks have been studied in  \cite{Secureconnectivity}-\cite{Geraci2014Physical}, respectively.
The impact of AN on the security of random networks has been studied in \cite{Enhancing}.

However, all of the above works focus on conventional microwave networks, and the obtained results can not be applied to mmWave networks directly,  due to the distinctive features of mmWave channel characteristics. For example, mmWave signals are more sensitive to blockage effects, and the fading statistical characteristics of the LOS link and NLOS link are totally different \cite{Millimeterpotentials}.
For characterizing the blockage effects of mmWave signals, different mmWave channel models have been proposed in \cite{CoverageMillimeter}-\cite{MultiTierMillimiterWave}. In \cite{AnalysisBlockage}, an exponential blockage model has been proposed, and such model has been approximated as a LOS ball based blockage model for the coverage analysis
in \cite{CoverageMillimeter,Coverageandcapacity}. In \cite{MillimeterAdhoc}, the authors adopted the exponential blockage model to perform the coverage and capacity analysis for mmWave ad hoc networks. In \cite{TractableModelMillimiterWave}, the authors proposed a ball based blockage model which is validated by using field measurements in New York and Chicago.
Taking the outage state emerging in the mmWave communication into consideration, in \cite{MultiTierMillimiterWave}, the authors have proposed a two-ball approximate blockage model for the analysis of the coverage and average rate of the multi-tier mmWave cellular network.

Under these new characteristics of mmWave channels, the secrecy performance of a mmWave cellular network will be significantly different from the conventional microwave network, which should be re-evaluated. The efficiency of traditional physical layer security techniques should be re-checked as well.
Recently, the secrecy performance of a point-to-point mmWave communication has been studied in \cite{SecureMillimeter}, which has shown that mmWave systems can enable significant secrecy improvement compared with conventional  microwave systems. However, the \emph{network-wide} secrecy performance of the mmWave cellular communication is still unknown, which motivates our work.

\subsection{Contribution}
In this paper, using the stochastic geometry framework and the blockage model proposed in \cite{TractableModelMillimiterWave}, we proposed a systematic secrecy performance analysis approach for the mmWave cellular communication, by modeling the random locations
of the BSs and eavesdroppers as two independent
homogeneous Possion point processes (PPPs).
Our contributions can be summarized as follows:
%Just as \cite{CoverageMillimeter}-\cite{MultiTierMillimiterWave}, we incorporate directional beamforming by approximating the actual beamforming pattern as a sectored model. Just as \cite{CoverageMillimeter,MillimeterAdhoc}, we approximate the mmWave small fading power via a Gamma random variable.
%Just as \cite{TractableModelMillimiterWave,MultiTierMillimiterWave}, we first consider the noise-limited mmWave cellular communication  which is applicable to medium/sparse network deployments.
%Assuming BSs adopt Wyner code to fight against eavesdropping,
%As indicated by \cite{CoverageMillimeter,MillimeterAdhoc}, the network interference power may dominate the noise power for the dense network.
%Therefore, just as \cite{CoverageMillimeter,MillimeterAdhoc}, taking the network interference into consideration, we characterize the average number of perfect communication links per unit area of the mmWave communication link in the presence of the non-colluding eavesdroppers.
\begin{enumerate}
\item \textbf{Secrecy performance analysis of noise-limited mmWave cellular networks.} We characterize the secrecy performance of a noise-limited mmWave cellular network that is applicable to medium/sparse network deployments, where each BS only adopts the directional beamforming to transmit the confidential information. Considering two cases: the non-colluding eavesdropper case and the colluding eavesdroppers case, we derive the analysis result of the secure connectivity probability  and the cumulative distribution function (CDF) of the received SNR at the typical receiver and eavesdropper, respectively.
    The secure connectivity probability facilitates the evaluation of the probability of the existence of secure connections from a typical transmitter to its intended receiver. With the CDF of the received SNR at the typical receiver and eavesdropper, we can characterize the average number of perfect communication links per unit area statistically in the random network.
    We show that the high gain narrow beam antenna is very important for enhancing the secrecy performance of mmWave networks.
    %This can be explained by the fact that the highly directional antenna array can reduce the information leakage to eavesdroppers and improve the signal receiving quality of the intended receiver. Furthermore, the length of the serving link is reducing with the increasing the intensity of BSs, which would improve the signal receiving quality of the intended receiver.
\item \textbf{Secrecy performance analysis of AN assisted mmWave cellular networks.}
 When AN is transmitted concurrently with the confidential information for
 interfering potential eavesdroppers, the AN radiation would increase the network interference. Thus, taking the network interference into consideration, we characterize  the CDF of  received SINRs at the intended receiver and non-colluding eavesdroppers. The secrecy probability and average number of perfect communication links per unit area for the AN assisted transmission have also been derived. The optimal power allocation between the AN and confidential signal is shown to depend on the array pattern and the intensity of eavesdroppers.
\end{enumerate}

\subsection{Paper Organization and Notations}
In Section II, the system model and mmWave channel characteristics are introduced. In Section III, considering the noise-limited mmWave cellular communication, we characterize the secure connectivity probability and  average number of perfect communication links per unit area. In Section IV, taking the inter-cell interference into consideration,
we characterize the average number of perfect communication links per unit area of the AN-assisted mmWave communication.
Numerical results are provided in Section V and the paper is concluded in Section VI.
%$b(o,D)$ denotes a circle whose radius is $D$ and center is origin.

\emph{Notation:} $x\sim\textrm{gamma}(k,m)$ denotes the gamma-distributed random variable with shape $k$ and scale $m$, $\gamma(x,y)$ is the lower incomplete gamma function \cite[8.350.1]{Table}, $\Gamma(x)$ is the gamma function \cite[eq. (8.310)]{Table}, and $\Gamma(a,x)$ is the upper incomplete function \cite[8.350.2]{Table}. $b(o,D)$ denotes the ball whose center is origin and radius is $D$.
 The factorial of a non-negative integer $n$ is denoted by $n!$,  $\mathbf{x} \sim \mathcal{CN}\left(\mathbf{\Lambda}, \mathbf{\Delta}\right)$ denotes the circular symmetric complex Gaussian vector with mean vector $\mathbf{\Lambda}$ and covariance matrix $\mathbf{\Delta}$, $\binom{n}{k}=\frac{n!}{\left(n-k\right)!k!}$. $\mathcal{L}_X(s)$ denotes the Laplace transform of $X$, i.e., $\mathbb{E}\left(e^{-sX}\right)$.
$_2F_1(\alpha,\beta;\gamma,z)$ is the Gauss hypergeometric function \cite[eq. (9.100)]{Table}.

%Besides the common notations mentioned above, the notations are summarized
%in Table I.

%\begin{table}[!tbp]
%\centering
%\caption{List of some important notations}
%\begin{tabular}{| l | p{6cm} |}
%\hline
%${\Phi}_B$ & Poisson point process (PPP) of  mmWave BSs,  \\
%${\Phi}_{B_L}$ & PPP of the LOS mmWave BSs,  \\
%${\Phi}_{B_N}$ & PPP of the NLOS mmWave BSs,  \\
%${\Phi}_{E}$ & PPP of the eavesdroppers,  \\
%$\Phi_I$&PPP of the interfere transmitting the confidential signals to the typical UE\\
%$\Phi_A$&PPP of the interfere transmitting the artificial noise to the typical UE\\
%$\lambda_B, \lambda_{B_L}, \lambda_{B_N}, \lambda_{E}$ & Densities of ${\Phi}_B, {\Phi}_{B_L}, {\Phi}_{B_N}, {\Phi}_{E}$, \\
%$C_S$ & Achievable ergodic secrecy rate,\\
%$r_L,r_N$& Distance between the UE and its serving BS in $\Phi_L$ and $\Phi_N$,\\
%$A_L$& Probability that the user  is associated with a LOS BS
%\\
%$A_N$& Probability that the user  is associated with a NLOS BS\\
%$\tau_n$& Secure connectivity probability in the presence of non-colluding eavesdroppers\\
%$\tau_c$& Secure connectivity probability in the presence of colluding eavesdroppers\\
%$N_p$& Average number of perfect communication links per unit\\
%\hline
%\end{tabular}
%\end{table}

\section{System Model and Problem Formulation}
We consider the downlink secure communication in the mmWave cellular network, where multiple spatially distributed BSs transmit the confidential information to authorized users  in the presence of multiple malicious eavesdroppers. In the following subsections, we first introduce the system model and channel characteristics adopted in this paper, which have been validated in \cite{CoverageMillimeter,MillimeterAdhoc,TractableModelMillimiterWave}. With such models, we give some important results on probability theory which will
be used in the performance analysis. The secrecy performance metrics adopted are given in Section II-G.

\subsection{BS and eavesdropper layout}
The locations of the BSs are modeled by a homogeneous PPP $\Phi_B$ of intensity $\lambda_B$.  Using PPP for modeling the irregular BSs locations has been shown to be an accurate and tractable approach for characterizing the downlink performance of the cellular network \cite{wirelessnetworks}. Just as \cite{Enhancing}-\cite{D2Dsecure}, the locations of multiple eavesdroppers are modeled  as an independent homogeneous PPP, $\Phi_E$, of intensity $\lambda_E$.
Such random PPP model is well motivated by the random and unpredictable eavesdroppers' locations.  Furthermore, just as \cite{Enhancing}, \cite{worstcase1}-\cite{worstcase3}, we consider the \textbf{worst-case} scenario  by
facilitating the eavesdroppers' multi-user decodability, i.e., eavesdroppers can perform successive interference cancellation \cite{FundamentalsWirelessCommunication} to eliminate the interference due to the information signals from other interfering BSs. The total transmit power of each BS is $P_t$.

%For compensating significant channel attenuation, the deployment of mmWave BSs becomes dense and irregular \cite{Millimeterpotentials}.
%Furthermore, just as \cite{CoverageMillimeter,MillimeterAdhoc,TractableModelMillimiterWave}, we assume that all mmWave BSs are fully-loaded (i.e., they have always data to transmit).
%\footnote{The core concept of the successive interference cancellation can be described as follows. Once the eavesdropper decodes the information signals transmitted from interfering BSs, it can reconstruct the information signals and subtract them for the aggregate received signal. Then, the eavesdropper can decode the information of interest in the absence of the interference from the information signals.  The details about the successive interference cancellation can be found in \cite[Charpter 6.1]{FundamentalsWirelessCommunication}. }

\subsection{Directional beamforming}
For compensating the significant path-loss at mmWave frequencies, highly directional beamforming antenna arrays are deployed at BSs to perform the directional beamforming. For mathematical tractability and similar to \cite{CoverageMillimeter,MillimeterAdhoc,TractableModelMillimiterWave,MultiTierMillimiterWave}, the antenna pattern is approximated by a sectored antenna model in  \cite{directionalAntenna}. In particular,
\begin{align}
G_b(\theta)=\left\{
\begin{array}{ll} M_s,&\mathrm{if}\quad |\theta|\leq\theta_b
\\ m_s,
&\textrm{Otherwise},
\end{array}
\right.\label{SectorAntennaModel}
\end{align}
where $\theta_b$ is the beam width of the main lobe, $M_s$ and $m_s$ are the array gains of main and sidelobes, respectively.
In this paper, we assume that each BS can get the perfect CSI estimation, including angles of arrivals and fading, and then, they can adjust their antenna steering orientation array for
adjusting the boresight direction of antennas to their intended receivers and
maximizing the directivity gains. In the following, we denote the boresight direction of the antennas as $0^o$.
Therefore, the directivity gain for the intended link is $M_s$. For each interfering link, the angle $\theta$ is independently and uniformly distributed in $\left[-\pi,\pi\right]$, which results in a random directivity gain $G_b(\theta)$. For simplifying the performance analysis,
just as \cite{TractableModelMillimiterWave,SecureMillimeter}, the authorized users and malicious eavesdroppers are both assumed to be  equipped with a single omnidirectional antenna in this paper. \footnote{This assumption is just for simplifying the performance analysis.  However, the obtained analysis methods can be extended to the multiple antennas case directly by modeling the array pattern at authorized users and malicious eavesdroppers in a similar way as (\ref{SectorAntennaModel}).}

\subsection{Small-scale fading}
Just as \cite{CoverageMillimeter,MillimeterAdhoc}, we assume that the small-scale fading of each link follows independent Nakagami fading, and the Nakagami fading parameter of the LOS (NLOS) link is $N_L$ ($N_N$). For simplicity, $N_L$ and $N_N$ are both assumed to be positive integers.
%Measurement results in \cite{5GMillimeter} show that highly directional antennas alleviate the effect of small-scale fading, especially for the LOS link. Such scenario can be approximated by using a large Nakagami fading parameter \cite{CoverageMillimeter,MillimeterAdhoc}.
In the following, the small-scale channel gain from the BS at $x\in\mathbb{R}^2$ to the authorized user (eavesdropper) at $y\in\mathbb{R}^2$ is expressed as $h_{xy}$ ($g_{xy}$).

\subsection{Blockage Model}
The blockage model proposed in \cite{TractableModelMillimiterWave} is adopted, which can be regarded as an approximation of  the statistical blockage model in \cite[eq. (8)]{MillimeterChannelModeling}, \cite{MultiTierMillimiterWave}, and incorporates the LOS ball model proposed in \cite{CoverageMillimeter,Coverageandcapacity} as a special case. As shown by  \cite{RateTrendsMillimiterWave,TractableModelMillimiterWave}, the blockage model proposed in \cite{TractableModelMillimiterWave} is simple yet flexible enough to capture blockage statistics, coverage and rate trends in mmWave cellular networks. In particular, defining $q_L(r)$ as the probability that a link of length $r$ is LOS,
\begin{align}
q_L(r)=\left\{
\begin{array}{ll} C,&\mathrm{if}\quad r\leq D,
\\ 0,
&\textrm{Otherwise},
\end{array}
\right.\label{BlockageModel}
\end{align}
for some $0\leq C\leq 1$. The parameter $C$ can be interpreted as the average LOS area in the spherical region around a typical user.
The empirical $(C,D)$ for Chicago and Manhattan are $(0.081,250)$ and $(0.117,200)$, respectively \cite{TractableModelMillimiterWave}, which would be adopted in the simulation results.
With such blockage model, the BS process in $b(o,D)$ can be divided into two independent PPPs: the LOS BS process $\Phi_{L}$ with intensity $C\lambda_B$ and NLOS BS process with intensity $(1-C)\lambda_B$ \cite[Proposition 1.3.5]{wirelessnetworks}. Outside $b(o,D)$, only the NLOS BS process exists with intensity $\lambda_B$. We denote the whole NLOS BS process as $\Phi_{N}$.

\subsection{Path loss model}
Just as \cite{CoverageMillimeter,TractableModelMillimiterWave}, different path loss laws are applied to LOS and NLOS links. In particular, given a link from $x\in\mathbb{R}^2$ to $y\in\mathbb{R}^2$, its path loss $L(x,y)$ can be calculated by
\begin{align}
L(x,y)=\left\{
\begin{array}{ll} C_L||x-y||^{-\alpha_L},&\textrm{ if link } x\rightarrow y\textrm{ is LOS link},
\\
C_N||x-y||^{-\alpha_N},
&\textrm{ if link } x\rightarrow y\textrm{ is NLOS link},
\end{array}
\right.\label{BlockageModel}
\end{align}
 where $\alpha_L$ and $\alpha_N$ are the LOS and NLOS path loss exponents, and $C_L\triangleq10^{-\frac{\beta_L}{10}}$ and $C_N\triangleq10^{-\frac{\beta_N}{10}}$ can be regarded as the path-loss intercepts  of LOS and NLOS links at the reference distance.
 Typical $\alpha_j$ and $\beta_j$ for $j\in\{L,N\}$ are defined in \cite[Table I]{MillimeterChannelModeling}. For exmple, for 28 GHz bands, $\beta_L=61.4$, $\alpha_L=2,$ and $\beta_N=72$, $\alpha_N=2.92$.
% For building a friendly mathematical model and simplifying the analysis, just as \cite{CoverageMillimeter,Coverageandcapacity,MillimeterAdhoc}, we do not consider the effect of the large-scale shadowing. The analysis involving the shadowing effects is left for the future work.
From the measured values of $C_j$ and $\alpha_j,$ $j\in\{L,N\}$ in \cite[Table I]{MillimeterChannelModeling}, we know that it satisfies $C_L>C_N$ and $\alpha_L<\alpha_N$.
%, we perform the millimeter network performance analysis on condition that $C_L>C_N$ and $\alpha_L<\alpha_N$. This not only simplifies the theoretical analysis, but also conforms to the actual conditions.

\subsection{User association}
For maximizing the receiving quality of authorized users \cite{CoverageMillimeter,TractableModelMillimiterWave}, one authorized user is assumed to be associated with the BS offering the lowest path loss to him, since the network considered is homogeneous. Thus, for the typical authorized user at origin, its serving BS is located at $x^*\triangleq\arg\max_{x\in\Phi_B}L(x,o)$.
Denoting the distance from the typical authorized user to the nearest BS in $\Phi_j$ as $d^*_{j}$ for $j=\{L,N\}$, the following Lemma 1 provides their probability distribution functions (pdf), and the obtained statistics hold for a generic authorized user, due to Slivnyak's theorem \cite{wirelessnetworks}.
\begin{lemma}
Given the typical authorized user observes at least one LOS BS, the pdf of $d^*_{L}$ is
\begin{align}
f_{d^*_{L}}(r)=\frac{2\pi C\lambda_B r\textrm{exp}(-\pi C\lambda_B r^2)}{1-\textrm{exp}(-\pi C\lambda_B D^2)},\textrm{ for } r\in\left[0,D\right].\label{pdfdl}
\end{align}
On the other hand, the pdf of $d^*_{N}$ is given by
\begin{align}
f_{d^*_N}(r)=2\pi (1-C)\lambda_Bre^{-\pi(1-C)\lambda_Br^2}\mathbb{I}\left(r\leq D\right)+2\lambda_B\pi re^{-\lambda_B\pi\left(r^2-D^2\right)}e^{-\pi(1-C)\lambda_BD^2}\mathbb{I}\left(r> D\right),
\label{distancedistribution}
\end{align}
where $\mathbb{I}\left(.\right)$ is the indicator function.
\end{lemma}
\begin{IEEEproof}
The proof is given in Appendix A.
\end{IEEEproof}
Then, the following lemma gives the probability that the typical authorized user is associated with a LOS or NLOS BS.
\begin{lemma}
The probability that the authorized user  is associated with a NLOS BS, $A_N$, is given by
\begin{align}
A_N=&\int^{\mu}_0{\left(\textrm{e}^{-\pi C\lambda_B\left(\frac{C_L}{C_N}\right)^{\frac{2}{\alpha_L}}x^{\frac{2\alpha_N}{\alpha_L}}}-\textrm{e}^{-\pi C\lambda_B D^2}\right)2\pi (1-C)\lambda_Bx}
\textrm{e}^{-\pi (1-C)\lambda_Bx^2}dx+\textrm{e}^{-\pi C\lambda_BD^2},\label{AN}
\end{align}
where $\mu\triangleq\left(\frac{C_L}{C_N}\right)^{-\frac{1}{\alpha_N}}D^{\frac{\alpha_L}{\alpha_N}}$.
The probability that the typical authorized user  is associated with a LOS BS is given by $A_L=1-A_N$.
\end{lemma}
\begin{IEEEproof}
The proof is given in Appendix B.
\end{IEEEproof}

With the smallest path loss association rule, the typical authorized user would be associated with the nearest LOS BS in $\Phi_L$ or the nearest NLOS BS in $\Phi_N$. The following lemma gives the pdf of the distance between the typical authorized user and its serving BS in $\Phi_j$, i.e., $r_j$, $\forall j\in\{L,N\}$.
%On the condition that the serving BS is in $\Phi_i$, for $i\in\{L,N\}$, the distance between the typical user and its serving BS is denoted as $r_i$ whose pdf is given by the following lemma.
\begin{lemma}
On the condition that the serving BS is in $\Phi_L$, the pdf of the distance from the typical authorized user to its serving BS in $\Phi_L$ is
\begin{align}
f_{r_L}(r)=\frac{\textrm{exp}\left(-(1-C)\lambda_B\pi\left(\frac{C_N}{C_L}\right)^{\frac{2}{\alpha_N}}r^{\frac{2\alpha_L}{\alpha_N}}\right)2\pi C\lambda_Br\textrm{exp}\left(-C\lambda_B\pi r^2\right)}{A_L}\textrm{,  } r\in\left[0,D\right].\label{frL}
\end{align}
On the condition that the serving BS is in $\Phi_N$, the pdf of the distance from the typical authorized user to its serving BS in $\Phi_N$ is
\begin{align}
&f_{r_N}(r)=\frac{2\pi\lambda_Br\textrm{exp}\left(-\pi\lambda_Br^2\right)\left(\left(1-C\right)\textrm{exp}\left(\pi C\lambda_B\left(r^2-D^2\right)\right)\mathbb{I}\left(r\leq D\right)+{\mathbb{I}\left(r\geq D\right)}\right)}{A_N}
%+\frac{2\pi\lambda_Br\textrm{exp}\left(-\pi\lambda_Br^2\right)\mathbb{I}\left(r\geq D\right)}{A_N}
\nonumber\\
&+\frac{2\pi (1-C)\lambda_Br e^{-(1-C)\lambda_B\pi r^2}\left(e^{-C\lambda_B\pi\left(\frac{C_L}{C_N}\right)^{\frac{2}{\alpha_L}}r^{\frac{2\alpha_N}{\alpha_L}}}-e^{-C\lambda_B\pi D^2}\right)}{A_N}\mathbb{I}\left(r\leq\left(\frac{C_N}{C_L}\right)^{\frac{1}{\alpha_N}}D^{\frac{\alpha_L}{\alpha_N}}\right).\label{frN}
\end{align}
\end{lemma}
\begin{IEEEproof}
The proof is given in Appendix C.
\end{IEEEproof}

\subsection{Secrecy Performance Metric}
In this paper, we assume that the channels are all quasi-static fading
channels. The legitimate receivers and eavesdroppers can obtain their own CSI, but
mmWave BSs do not know the instantaneous CSI of eavesdroppers. For protecting the confidential information from wiretapping, each BS encodes the confidential data by the Wyner code \cite{Wyner}. Then two code rates
namely, the rate of the transmitted codewords $R_b$,
and the rate of the confidential information $R_s$ should be determined before the data transmission, and $R_b-R_s$ is the cost for securing the confidential information. The details of the code construction can be found in \cite{Wyner,NewPhysicalLayerSecurity}. In this paper, just as \cite{Enhancing, D2Dsecure, NewPhysicalLayerSecurity}, we adopt the fixed rate transmission, where
$R_b$ and $R_s$ are fixed during the information transmission.  For the secrecy transmission over quasi-static fading
channels, the perfect secrecy can not always be guaranteed. Therefore, as indicated in \cite{Secureconnectivity, D2Dsecure, NewPhysicalLayerSecurity}, an outage-based secrecy performance metric is more suitable. Therefore, we analyze the secrecy performance of the mmWave  communication by considering both the secure connectivity probability  and average number of perfect communication links per unit area.

\begin{enumerate}
\item {\emph{Secure connectivity probability \cite{Secureconnectivity}}}. Secure connectivity probability introduced in \cite{Secureconnectivity}, is defined as the probability that the  secrecy rate is nonnegative.
Using the secure connectivity probability, we aim to statistically characterize the existence of secure connection between any randomly chosen BS and its intended authorized user in the presence of multiple  eavesdroppers.
\item \emph{Average number of perfect communication links per unit area  \cite{D2Dsecure}}.When $R_b$ and $R_s$ are given, we define the links that have perfect connection and secrecy as perfect communication links. Then, the mathematical definition of the average number of  perfect communication links per unit area is given as follows.
\begin{itemize}
\item \textbf{Connection Probability.}
When $R_b$ is below the capacity of legitimate links,  authorized users can decode  signals with an arbitrary small error, and thus perfect connection can be assured. Otherwise, connection outage would occur.
%Defining $R_C\Triangleq\Textrm{Log}_2\Left(1+T_C\Right)$ And $T_C$ Is The Threshold For Connection,
The connection probability is denoted as $p_{con}$.
\item\textbf{{Secrecy Probability.}}
When the wiretapping capacity of eavesdroppers is below the rate redundancy $R_e\triangleq R_b-R_s$, there will be no information leakage to potential eavesdroppers, and thus perfect secrecy of the link can be assured \cite{Wyner}. Otherwise, secrecy outage would occur.
%Defining $R_e\triangleq\textrm{log}_2\left(1+T_e\right)$ and $T_e$ is the  threshold for secrecy transmission.
The secrecy probability is denoted as $p_{sec}$.
%The secrecy probability can be calculated by
%\begin{align}
%p_{sec}=\textrm{Pr}\left(\textrm{SINR}_{E}\leq T_e\right),
%\end{align}
%where SINR$_E$ is the received SINR at the most detrimental eavesdropper in $\Phi_E$.
\end{itemize}
Following \cite[eq. (29)]{D2Dsecure}, the average number of perfect communication links per unit area is
\begin{align}
N_p = \lambda_Bp_{con}p_{sec}. \label{PerfectCommunicationLink}
\end{align}
\end{enumerate}
\begin{remark}
With the given $R_b$ and $R_s$, the average achievable secrecy throughput per unit area $\omega$ can be calculated by
$
\omega=N_pR_s.
$
\end{remark}

\section{Secrecy Performance of the Noise-Limited Millimeter Wave Communication}
In this section, we evaluate the secrecy performance of the direct transmission for the noise-limited mmWave communication.
As pointed out by \cite{Millimeterpotentials,MillimeterChannelModeling,TractableModelMillimiterWave,MultiTierMillimiterWave}, highly directional transmissions used in mmWave systems combined with short cell radius results in links that are noise-dominated, especially for densely blocked settings (e.g., urban settings) and medium/sparse network deployments \cite{TractableModelMillimiterWave,MultiTierMillimiterWave}. This distinguishes from  current dense cellular deployments where links are overwhelmingly
interference-dominated. Therefore, just as \cite{TractableModelMillimiterWave,MultiTierMillimiterWave}, we first study the secrecy performance of the noise-limited mmWave communication without considering the effect of inter-cell interference. The received SNR by the typical authorized user at origin and the eavesdropper at $z$ with respect to the serving BS can be expressed as $\textrm{SNR}_U=\frac{P_tM_sL(x^*,o)h_{x^*o}}{N_0}$ and $\textrm{SNR}_{E_z}=\frac{P_tG_b(\theta)L(x^*,z)g_{x^*z}}{N_0}$. $N_0$ is the noise power in the form of
$N_0=10^{\frac{N_0(dB)}{10}}$, where $N_0(dB)=-174+10\textrm{log}_{10}(\textrm{BW})+\mathcal{F}_{dB}$, BW is the transmission bandwidth and  $\mathcal{F}_{dB}$ is the noise figure \cite{MultiTierMillimiterWave}.
 With the array pattern in (\ref{SectorAntennaModel}), $G_b(\theta)$ seen by the eavesdropper is a Bernoulli  random variable whose probability mass function (PMF) is given by
\begin{align}
G_b(\theta)=\left\{\begin{matrix}
M_s,& \textrm{Pr}_{G_b}(M_s)\triangleq \textrm{Pr}\left(G_b(\theta)=M_s\right)= \frac{\theta_b}{180},
\\
m_s,&\ \ \textrm{Pr}_{G_b}(m_s) \triangleq \textrm{Pr}\left(G_b(\theta)=m_s\right)=\frac{180-\theta_b}{180}.
\end{matrix}
\right.
%&p_{G_b(\theta)}(M_s)=\textrm{Pr}\left(G_b(\theta)=M_s\right)=\frac{\theta_b}{180},
%\quad p_{G_b(\theta)}(m_s)=\textrm{Pr}\left(G_b(\theta)=m_s\right)=\frac{180-\theta_b}{180}.
\end{align}

\subsection{Non-colluding Eavesdroppers}
In this subsection, assuming that the random distributed eavesdroppers are \textbf{non-colluding}, we evaluate the secrecy performance of the mmWave cellular network.

\subsubsection{Secure Connectivity Probability}
We first study the secure connectivity probability, $\tau_n$, of the mmWave communication in the presence of multiple {non-colluding eavesdroppers}.
A secure connection is possible if the condition $\frac{M_sL(x^*,o)h_{x^*o}}{\max_{z\in\Phi_E}G_b(\theta)L(x^*,z)g_{x^*z}}\geq 1$ holds \cite{Secureconnectivity}, and the secure connectivity probability can be calculated by $\tau_n=\textrm{Pr}\left(\frac{M_sL(x^*,o)h_{x^*o}}{\max_{z\in\Phi_E}G_b(\theta)L(x^*,z)g_{x^*z}}\geq 1\right)$. We can see that the wiretapping capability of multiple eavesdroppers is determined by the path loss process $G_b(\theta)L(x^*,z)g_{x^*z}$. Thus, for facilitating the performance evaluation, the following process is introduced.
\begin{definition}

The path loss process with fading (PLPF), denoted as $\mathcal{N}_E$, is the point process on $\mathbb{R}^+$ mapped from $\Phi_E$, where $\mathcal{N}_E\triangleq\left\{\varsigma_z=\frac{1}{G_b(\theta)g_{xz}L(x,z)},z\in\Phi_E\right\}$ and $x$ denotes the location of the wiretapped BS.
We sort the elements of $\mathcal{N}_E$ in ascending order and denote the sorted elements of $\mathcal{N}_E$ as $\left\{\xi_i,i=1,\ldots\right\}$.
The index is introduced such that $\xi_i\leq \xi_j$ for $\forall i<j$.
\end{definition}

Note that $\mathcal{N}_E$ involves both the impact of small fading and spatial distribution of eavesdroppers, which is an ordered process. Consequently, $\mathcal{N}_E$ determines the wiretapping capability of eavesdropper. We then have the following lemma.
\begin{lemma}
The PLPF $\mathcal{N}_E$ is an one-dimensional nonhomogeneous PPP with the intensity measure
\begin{align}
\Lambda_E\left(0,t\right)=&2\pi\lambda_E\left(\sum_{j\in\left\{L,N\right\}}q_j\left(\Omega_{j,\textrm{in}}\left(M_s,t\right)+\Omega_{j,\textrm{in}}\left(m_s,t\right)\right)+\Omega_{N,\textrm{out}}\left(M_s,,t\right)
+\Omega_{N,\textrm{out}}\left(m_s,t\right)\right),\label{LambdaE}
\end{align}
where $q_L\triangleq C$, $q_N\triangleq 1-C$, and
$\Omega_{j,\textrm{in}}(V,t) \triangleq \textrm{Pr}_{G_b}(V)\frac{\left(VC_jt\right)^{\frac{2}{\alpha_j}}}{\alpha_j}\sum^{N_j-1}_{m=0}\frac{\gamma\left(m+\frac{2}{\alpha_j},\frac{D^{\alpha_j}}{VC_jt}\right)}{m!}
$,
$\Omega_{j,\textrm{out}}(V,t) \triangleq \textrm{Pr}_{G_b}(V)\frac{\left(VC_jt\right)^{\frac{2}{\alpha_j}}}{\alpha_j}\sum^{N_j-1}_{m=0}\frac{\Gamma\left(m+\frac{2}{\alpha_j},\frac{D^{\alpha_j}}{VC_jt}\right)}{m!}
$, with $V\in\left\{M_s,m_s\right\}$.
\end{lemma}
\begin{IEEEproof}
The proof is given in Appendix D.
\end{IEEEproof}

The following theorem gives the analysis result of the secure connectivity probability in the presence of non-colluding eavesdroppers.
\begin{Th}
In the case of non-colluding eavesdroppers, the secure connectivity probability is
\begin{align}
\tau_n=\sum_{j\in\{L,N\}}
A_j\int^D_0f_{r_j}(r)dr\int^{+\infty}_0\frac{\textrm{e}^{-\Lambda_E\left(0,\frac{r^{\alpha_j}}{M_swC_j}\right)}w^{N_j-1}e^{-w}}{\Gamma(N_j)}dw.
\label{SecureConnectivityProbability}
\end{align}
\end{Th}
\begin{IEEEproof}
We have the following derivations:
\begin{align}
\textrm{Pr}&\left(\frac{M_sL(x^*,o)h_{x^*o}}{\max_{z\in\Phi_E}G_b(\theta)L(x^*,z)g_{x^*z}}\geq 1\right)=
\textrm{Pr}\left(\min_{z\in\Phi_E}\frac{1}{G_b(\theta)L(x^*,z)g_{x^*z}}\geq \frac{1}{M_sL(x^*,o)h_{x^*o}}\right)
\nonumber\\
&\overset{(e)}{=}\textrm{Pr}\left(\xi_1\geq \frac{1}{M_sL(x^*,o)h_{x^*o}}\right)\overset{(f)}{=}\mathbb{E}_{L(x^*,o),h_{x^*o}}\left(\textrm{exp}\left(-\Lambda_E\left(0,\frac{1}{M_sL(x^*,o)h_{x^*o}}\right)\right)\right)
\nonumber\\
&\overset{(g)}{=}A_L\mathbb{E}_{r_L,h_{x^*o}}\left(\textrm{exp}\left(-\Lambda_E\left(0,\frac{r_L^{\alpha_L}}{M_sC_Lh_{x^*o}}\right)|\textrm{Serving BS is a LOS BS}\right)\right)+
\nonumber\\
&\qquad A_N\mathbb{E}_{r_N,h_{x^*o}}\left(\textrm{exp}\left(-\Lambda_E\left(0,\frac{r_N^{\alpha_N}}{M_sC_Nh_{x^*o}}\right)|\textrm{Serving BS is a NLOS  BS}\right)\right),\label{derivationofsecureconnectivity}
\end{align}
where step $(e)$ is due to Definition 1, step $(f)$ follows the PPP's void probability \cite{wirelessnetworks}, and step ($g$) is due to the law of total probability. When the serving BS  is a LOS BS, $h_{x^*o}\sim\textrm{gamma}\left(N_L,1\right)$ and the pdf of $r_L$ is given by (\ref{frL}), and  when the serving BS  is a NLOS BS, $h_{x^*o}\sim\textrm{gamma}\left(N_N,1\right)$ and the pdf of $r_N$ is given by (\ref{frN}).
Finally, substituting the pdf of $h_{x^*o}$, $r_L$, and $r_N$ into (\ref{derivationofsecureconnectivity}), $\tau_n$ can be obtained.
\end{IEEEproof}
%\begin{remark}
%Although a  closed form of the analytical result of
%(\ref{SecureConnectivityProbability}) is difficult to get, the double integral term
%involved in (\ref{SecureConnectivityProbability}) can be evaluated with the iterative numerical method  in \cite{NumericalIntegration}, which facilitates the evaluation of the secure connectivity of mmWave communication in the presence of multiple non-colluding eavesdroppers.
%\end{remark}

\begin{figure}[!t]
\centering
\includegraphics[width=2.5in]{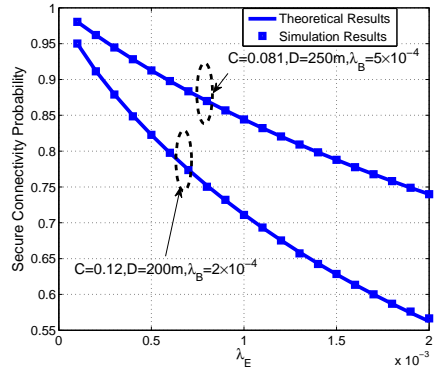}
\caption{Secure connectivity probability of the mmWave communication in the presence of multiple non-colluding eavesdroppers vs $\lambda_E$ with BW=2GHz, $P_t=30$ dB, $\mathcal{F}_{dB}=10$, $\theta_b=9^o$, $M_s=15$ dB, and $m_s=-3$ dB.}
\label{noncolludingEvessecureconnectivity}
\end{figure}
Theoretical results in Theorem 1 are validated in Fig. \ref{noncolludingEvessecureconnectivity}, where we plot the secure connectivity probability $\tau_{n}$ versus $\lambda_E$.
For all
the simulations in this paper, 100000 trials are used.
From Fig. \ref{noncolludingEvessecureconnectivity}, we can find that theoretical curves coincide with the simulation ones well, which validates the theoretical result in Theorem 1.

\subsubsection{Average number of perfect communication links per unit area}
In the following, we study the average number of perfect communication links per unit area, $N_p$, of the mmWave communication in the presence of \textbf{non-colluding} eavesdroppers.
Firstly, we should derive the analytical result of the connection probability and secrecy probability of a mmWave communication link, given by
\begin{align}
p_{con}\triangleq\textrm{Pr}\left(\textrm{SNR}_U\geq T_c\right)\textrm{ and }p_{sec,n}\triangleq\textrm{Pr}\left(\max_{z\in\Phi_E}\textrm{SNR}_{E_z}\leq T_e\right),
\end{align}
respectively, where $T_c\triangleq 2^{R_c-1}$ and $T_e\triangleq 2^{R_e-1}$. We have the following theorem.
\begin{Th}
For the non-colluding eavesdroppers case, the analytical result of $p_{con}$ is given by
\begin{align}
p_{con} = \int^{D}_0\frac{\Gamma\left(N_L,\frac{N_0T_cr^{\alpha_L}}{P_tM_SC_L}\right)}{\Gamma(N_L)}f_{r_L}(r)dr A_L+
\int^{+\infty}_0\frac{\Gamma\left(N_N,\frac{N_0T_cr^{\alpha_N}}{P_tM_SC_N}\right)}{\Gamma(N_N)}f_{r_N}(r)dr A_N,
\label{connectionnon}
\end{align}
and the analytical result of $p_{sec,n}$ is given by
\begin{align}
p_{sec,n}=\textrm{exp}\left(-\Lambda_E\left(0,\frac{1}{TeN_0}\right)\right).\label{Secrecynon}
\end{align}
\end{Th}
\begin{IEEEproof}
$p_{con}$ can be derived as follows
\begin{align}
p_{con}=&\textrm{Pr}\left(h_{x^*o}\geq \frac{N _0T_c}{P_tM_SL(x^*,o)}|\textrm{Serving BS is a LOS BS}\right)A_L
+\nonumber\\
&\textrm{Pr}\left(h_{x^*o}\geq \frac{N _0T_c}{P_tM_SL(x^*,o)}|\textrm{Serving BS is a NLOS BS}\right)A_N
\nonumber\\
=&\int^{D}_0\frac{\Gamma\left(N_L,\frac{N_0T_cr^{\alpha_L}}{P_tM_SC_L}\right)}{\Gamma(N_L)}f_{r_L}(r)drA_L+
\int^{+\infty}_0\frac{\Gamma\left(N_N,\frac{N_0T_cr^{\alpha_N}}{P_tM_SC_N}\right)}{\Gamma(N_N)}f_{r_N}(r)drA_N.
\end{align}
$p_{sec,n}$ can be derived as follows
\begin{align}
p_{sec,n}&=\textrm{Pr}\left\{\frac{\max_{z\phi_{E_z}}G_b(\theta)L(x^*,z)g_{x^*,z}}{N_0}\leq T_e\right\}\overset{(g)}{=}\textrm{Pr}\left\{\frac{1}{\xi_1N_0}\leq T_e\right\}\overset{(h)}{=}\textrm{exp}\left(-\Lambda_E\left(0,\frac{1}{T_e N_0}\right)\right),
\end{align}
where step $(g)$ is due to Definition 1, and step $(h)$ is due to the PPP's void probability \cite{wirelessnetworks}.
\end{IEEEproof}

\subsection{Colluding Eavesdroppers}
In this subsection, we study the secrecy performance of the mmWave communication by considering the worst case: \textbf{colluding eavesdroppers}, where distributed eavesdroppers adopt the maximal-ratio combining to process the wiretapped confidential information.
%Therefore, compared with the non-colluding eavesdroppers case, the secrecy performance of the mmWave communication would deteriorate.

\subsubsection{Secure Connectivity Probability}
The secure connectivity probability $\tau_c$ in the presence of multiple colluding eavesdroppers can be calculated by
\begin{align}
\tau_c=\textrm{Pr}\left(\frac{M_sL(x^*,o)h_{x^*o}}{I_E}\geq 1\right),
\end{align}
where $I_E\triangleq {\sum_{z\in\Phi_E}G_b(\theta)L(x^*,z)g_{x^*z}}$.
We have the following theorem.
\begin{Th}
In the case of colluding eavesdroppers, $\tau_{c}$ can be calculated by
\begin{align}
&\tau_{c} =\mathbb{E}_{r_j}\left[\sum_{j\in\{L,N\}}\sum^{N_{j}-1}_{m=0}\left(\frac{r_j^{\alpha_j}}{M_SC_j}\right)^m\frac{A_j}{\Gamma(m+1)}(-1)^m\mathcal{L}^{(m)}_{I_E}\left(\frac{r_j^{\alpha_j}}{M_SC_j}\right)\right],
\end{align}
where $\mathcal{L}_{I_E}(s)\triangleq\textrm{exp}\left(\Xi(s)\right)$ and
\begin{align}
\Xi(s)\triangleq&-s\left(2\pi\lambda_E\sum_{j\in\{L,N\}}q_j\left(\sum_{V\in\{M_s,m_s\}}\textrm{Pr}_{G_b}(V)\frac{\left(VC_j\right)^{\frac{2}{\alpha_j}}}{\alpha_j}\sum^{N_j-1}_{m=0}\frac{\left(D^{\alpha_j}/(VC_j)\right)^{m+\frac{2}{\alpha_j}}}{\left(m+\frac{2}{\alpha_j}\right)\left(s+{D^{\alpha_j}/(VC_j)}\right)^{m+1}}
\right.\right.
\nonumber\\
&\left.
{_2}F_1\left(1,m+1;m+\frac{2}{\alpha_j}+1;\frac{{D^{\alpha_j}/(VC_j)}}{{D^{\alpha_j}/(VC_j)}+s}\right)\Bigg)+
2\pi\lambda_E\sum_{V\in\{M_s,m_s\}}\textrm{Pr}_{G_b}(V)\frac{\left(VC_N\right)^{\frac{2}{\alpha_N}}}{\alpha_N}
\right.\nonumber\\
&\left.\sum^{N_N-1}_{m=0}
\frac{\left(D^{\alpha_N}/(VC_N)\right)^{m+\frac{2}{\alpha_N}}}{\left(1-\frac{2}{\alpha_N}\right)\left(s+D^{\alpha_N}/(VC_N)\right)^{m+1}}
{_2}F_1\left(1,m+1;2-\frac{2}{\alpha_N};\frac{s}{s+D^{\alpha_N}/(VC_N)}\right)
\right).
\end{align}
\end{Th}
\begin{IEEEproof}
The proof is given in Appendix E.
\end{IEEEproof}

Although the analytical result given in Theorem 3 is general and exact,
it is rather unwieldy, motivating the interest in acquiring a more compact
expression. Exploring the tight lower bound of the CDF of the gamma random variable in \cite{Inequality}, a tight upper bound of $\tau_{c}$ can be calculated as follows.

\begin{Th}
$\tau_c$ can be tightly upper bounded by
\begin{align}
&\tau_c\lessapprox
\sum_{j\in\{L,N\}}\sum^{N_j}_{n=1}\binom{N_j}{n}(-1)^{n+1}\int^{+\infty}_0f_{r_j}(r)\mathcal{L}_{I_E}\left(\frac{a_jnr^{\alpha_j}}{M_SC_j}\right)dr,
\end{align}
where $a_L\triangleq(N_L)^{-\frac{1}{N_L}}$ and $a_N\triangleq(N_N)^{-\frac{1}{N_N}}$.
\end{Th}
\begin{IEEEproof}
We leverage the tight lower bound of the CDF of  a normalized gamma random variable, $g$ with $N$ degrees of freedom as $\textrm{Pr}\left(x\leq y\right)\gtrapprox \left(1-e^{-\kappa y}\right)^N$  \cite{Inequality}, where $\kappa=\left(N!\right)^{-\frac{1}{N}}$. Since $h_{x^*,o}$ is a normalized gamma random variable, we have
\begin{align}
\tau_c\lessapprox 1-\sum_{j\in\{L,N\}}\mathbb{E}_{I_E,r_j}\left(\left(1-\textrm{exp}\left(-\frac{na_jI_Er_j^{\alpha_j}}{M_SC_j}\right)\right)^{N_j}\right).\label{Bound}
\end{align}
Using the binomial expansion, we can obtain
\begin{align}
\tau_c&\lessapprox \sum_{j\in\{L,N\}}\sum^{N_j}_{n=1}\binom{N_j}{n}(-1)^{n+1}\mathbb{E}_{r_j}\left(\mathbb{E}_{I_E}\left(\textrm{exp}\left(-\frac{a_jnr_j^{\alpha_j}I_E}{M_SC_j}\right)\right)\right)
\nonumber\\
&=\sum_{j\in\{L,N\}}\sum^{N_j}_{n=1}\binom{N_j}{n}(-1)^{n+1}\int^{+\infty}_0f_{r_j}(r)\mathcal{L}_{I_E}\left(\frac{a_jnr^{\alpha_j}}{M_SC_j}\right).
\end{align}
\end{IEEEproof}

\begin{figure}[!t]
\centering
\includegraphics[width=2.5in]{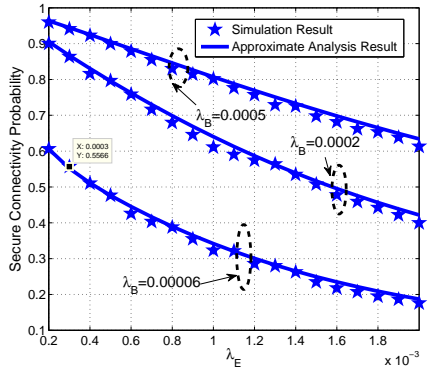}
\caption{Secure connectivity probability of mmWave communication in the presence of multiple colluding eavesdroppers vs $\lambda_E$. The system parameters are $P_t=30$ dB, $\beta_L=61.4$ dB, $\alpha_L = 2,$ $\beta_N=72$ dB, $\alpha_N=2.92$, BW = 2 GHz, $\mathcal{F}_{dB}=10$, $\lambda_B=0.0005,0.0002,0.00006,C=0.12$, $D=200$ m, $\theta_b=9^o$, $M_s=15$ dB, and $m_s=-3$ dB.}
\label{colludingEvessecureconnectivity}
\end{figure}

The bounds in Theorem 4 are validated in Fig. \ref{colludingEvessecureconnectivity}, where we plot the secure connectivity probability $\tau_{c}$ versus $\lambda_E$. From Fig. \ref{colludingEvessecureconnectivity}, we can find that theoretical curves coincide with simulation ones well, which show that the upper bound given in Theorem 4 is tight.

\subsubsection{Average number of perfect communication links per unit area}
In the case of colluding eavesdroppers, the connection probability $p_{con}$ of the typical authorized user can be still calculated by (\ref{connectionnon}) in Theorem 2, and the achievable secrecy probability $p_{sec}$ can be calculated by
\begin{align}
p_{sec,c}=\textrm{Pr}\left\{\frac{P_tI_E}{N_0}\leq T_e\right\}.\label{colludingEavesdropperpsec2}
\end{align}
For getting the analysis result of $p_{sec,c}$ in (\ref{colludingEavesdropperpsec2}), the CDF of $I_E$ should be available. Although the CDF of $I_E$ can be obtained from its Laplace transform $\mathcal{L}_{I_E}(s)$ by using the inverse Laplace transform calculation \cite{InverseLaplaceTransform}, it could get computationally intensive in certain cases and may render the analysis intractable. As an alternative, we resort to an approximation method widely adopted in \cite{CoverageMillimeter,MillimeterAdhoc,TractableModelMillimiterWave} for getting an approximation of $p_{sec,c}$ which is given in the following theorem.
\begin{Th}
In the case of  multiple colluding eavesdroppers, the approximation of $p_{sec,c}$ is given by
\begin{align}
 p_{sec,c}\lessapprox\sum^{N}_{n=1}\left(-1\right)^{n+1}\mathcal{L}_{I_E}\left(-\frac{an}{n_0T_e}\right),\label{SecrecyProbabilityBound}
\end{align}
where $\mathcal{L}_{I_E}(s)$ is given in Theorem 3, $a\triangleq(N!)^{\frac{1}{N}}$ and $N$ is the number of terms used in approximation.
\end{Th}
\begin{IEEEproof}
\begin{align}
p_{sec,c}=\textrm{Pr}\left\{\frac{P_tI_E}{N_0T_e}\leq 1\right\}\overset{(i)}{\approx}\textrm{Pr}\left\{\frac{P_tI_E}{N_0T_e}\leq w\right\}.
\end{align}
In $(i)$, $w$ is a normalized gamma random variable with a shape parameter, $N$, and the approximation in $(i)$ is due to the fact that a normalized gamma random variable converges to identity when its shape parameter goes to infinity  \cite{CoverageMillimeter,MillimeterAdhoc}.

Then, using the tight lower of the CDF of a normalized gamma random variable in \cite{Inequality}, $p_{sec}$ can be tightly upper bounded  by
\begin{align}
p_{sec,c}\lessapprox 1-\left(1-\textrm{exp}\left(-\frac{a P_tI_E}{N_0T_e}\right)\right)^N.\label{boundDerivation}
\end{align}
Finally, using  the binomial expansion, (\ref{boundDerivation}) can be further rewritten as (\ref{SecrecyProbabilityBound}).
\end{IEEEproof}

\begin{figure}[!t]
\centering
\includegraphics[width=2.5in]{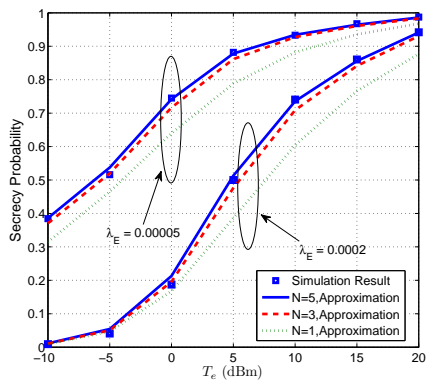}
\caption{Secrecy Probability of the mmWave communication in the presence of multiple colluding eavesdroppers vs $T_e$. The system parameters are $P_t=30$ dB, $\theta_b=9^o$, $M_s=15$ dB, $m_a=-3$ dB, $\beta_L=61.4$ dB, $\alpha_L=2,$ $\beta_N=72$ dB, $\alpha_N=2.92$, BW = 2 GHz, $\mathcal{F}_{dB}=10$, $\lambda_B=0.0005,C=0.081$, and $D=250$ m.}
\label{colludingEvessecureconnectivity2}
\end{figure}

The approximate analysis result in Theorem 5 is validated in Fig. \ref{colludingEvessecureconnectivity2}. From Fig. \ref{colludingEvessecureconnectivity2}, we can find that when $N=5,$ (\ref{SecrecyProbabilityBound}) can give an accurate approximation. Then, in the following simulations, we set $N=5$ to calculate  $p_{sec,c}$, approximately.

\section{Secrecy Performance  of the Interference-Limited mmWave Network with AN}
AN has been proved to be an efficient secure transmission strategy for the conventional cellular network \cite{Goel:TWC08,Enhancing}. But for the mmWave network, the utility of the AN should be re-evaluated due to the distinguishing features of the mmWave communication.
In this section, we will analyze the secrecy performance of the AN-assisted mmWave communication.  Since the additional AN would increase the network interference, different from the previous section,
we analyze the secrecy performance of the AN assisted mmWave communication by taking the inter-cell interference into consideration\footnote{In the mmWave network with AN, the transmitted AN simultaneously from each BS has made the transmissions of mmWave signals no longer highly directional. Therefore, different from mmWave network without AN, the out-cell interference should be taken into consideration  in the mmWave network with AN.}.
For obtaining a tractable problem, we only study the second secrecy performance metric: the average number of perfect communication links per unit area for the \textbf{non-colluding} eavesdroppers case.

%For the dense network with a large intensity of BSs, the typical authorized user would observe serval LOS interfering BSs, which makes the mmWave network be interference-limited \cite{CoverageMillimeter,MillimeterAdhoc}. Specially, in \cite{CoverageMillimeter}, the authors conclude that the mmWave network is ultra-dense when the average number of LOS BSs that a typical authorized user observes, $\rho>10$, where $\rho=C\lambda_B\pi D^2$ \cite[Theorem 4]{CoverageMillimeter}. Therefore,  we conclude that when
%$\lambda_B\geq \frac{10}{C\pi D^2}$, the mmWave network can be regarded as a dense mmWave network and the effect of the network interference can not be neglected \cite[Section IV]{CoverageMillimeter}.

By introducing different phase shifts in each directional antenna, each BS can concentrate the transmit power of the confidential information signals into the direction of its intended receiver, while radiating AN uniformly in all other directions. For tractability of the analysis, the actual array pattern of each BS is approximated by the model of sectoring with artificial noise proposed in \cite[Section II-A]{Enhancing}.
In particular, for the confidential information signals, it has main lobe of gain $M_s$ and angle of spread $\theta_b$, and just as \cite{Enhancing,MicrowaveCommunication}, the sidelobes of the confidential information signals are suppressed sufficiently, which can be omitted in the following\footnote{We should point out that the analysis results obtained in this section can be generalized to incorporate the
sidelobe leakage signals, by considering the eavesdroppers in the intended sector and outside the intended sector separately, just as Section III. However, the analysis results in such case would become more complicated, whilst few design insights can be brought. Furthermore, as we know, massive antenna array would be deployed at the mmWave BS for improving the mmWave signal transmission performance \cite{5GMillimeter, MillimeterChannelModeling}. Therefore, the sidelobes of the antenna pattern at  the mmWave BS can be suppressed sufficiently, and it is reasonable to omit the sidelobe in the  theoretical analysis.}. Accordingly, for the AN, it has main lobe of gain $M_a$ and angle of spread $360-\theta_b$, and the sidelobes of the AN are suppressed sufficiently, which can also be omitted. The sectors of the confidential signals and AN are non-overlapping.

Assuming that $\phi P_t$ is allocated to transmit the confidential information in the intended sector, and $(1-\phi) P_t$ is allocated to transmit AN concurrently out of intended sectors, where $0\leq \phi\leq 1$. The transmit power $x(\theta)$ of each BS is
\begin{align}
x(\theta){=}
\left\{
\begin{array}{lll}
M_s\phi P_t,&\mathrm{if}\quad |\theta|\leq \theta_b, &\textrm{Pr}_{x}(M_s)\triangleq \textrm{Pr}\left(x(\theta)=M_s\phi P_t\right) =\frac{\theta}{180},
\\ M_a(1-\phi)P_t,
&\textrm{Otherwise},&\textrm{Pr}_{x}(M_a)\triangleq\textrm{Pr}\left(x(\theta)=M_a(1-\phi)P_t\right) =\frac{180-\theta}{180}.
\end{array}
\right.
\label{SectorAntennaModelArtificialNoise}
\end{align}

%Specially,
%Then, with the sectorized antenna model in (\ref{SectorAntennaModel}) and (\ref{SectorAntennaModelArtificialNoise1}),  the transmitted signal $x(\theta)$ at each BS is
%\begin{align}
%x(\theta)=\left\{
%\begin{array}{ll} m_a(1-\phi)P_tn_a+M_s\phi P_ts,&\mathrm{if}\quad |\theta|\leq \theta_b,
%\\ M_a(1-\phi)P_tn_a+m_s\phi P_ts,
%&\textrm{Otherwise},
%\end{array}
%\right.\label{ApproximatePatternSignal}
%\end{align}
%where $n_a$ denotes the white Gaussian AN signal with variance 1. The sectored pattern in (\ref{ApproximatePatternSignal}) is an approximation of the  beam pattern of the signal and AN in \cite{ArtificialNoiseLinearAntenna}.
%Since the number antennas equipped at each BS is large, sidelobes are much lower compared with the main lobe. For facilitating the performance analysis in the following, just as \cite{Enhancing}, sidelobes are assumed to be suppressed sufficiently, and thus can be ignored.
%$x(\theta)$ can be further approximated as
%the approximation of the transmitted signal $x(\theta)$ is adopted in the following performance analysis. Similar approximation has been adopted in \cite{Enhancing} for evaluating the secrecy performance of the AN-assisted ad hoc network.

Then, according to the mapping theorem, for any receiver (authorized user or eavesdropper), the  interfering BSs can be divided into two independent PPPs: 1) the one transmitting the confidential signals to the receiver, which is denoted by $\Phi_I$ of intensity ${\lambda_B}\textrm{Pr}_{x}(M_s)$; 2) the one transmitting the AN to the receiver, which is denoted by $\Phi_A$ of intensity ${\lambda_B}\textrm{Pr}_{x}(M_a)$.

\subsection{Connection Probability}
Considering the typical authorized user at the origin, its received SINR$_U$ can be calculated by
\begin{align}
\textrm{SINR}_U=\frac{\phi P_tM_sh_{x^*o}L\left(x^*,o\right)}{I_B+N_0},
\end{align}
where the interference from multiple interfering BSs: $I_B\triangleq\sum_{y\in\Phi_I/{x^*}}M_s\phi P_th_{yx^*}L(y,x^*)+\sum_{y\in\Phi_A}M_a(1-\phi) P_th_{yx^*}L(y,x^*)$.
We have the following theorem.
\begin{Th}
The connection probability of the typical communication link can be tightly upper bounded by
\begin{align}
p_{con}\triangleq\textrm{Pr}\left(\textrm{SINR}_U\geq {T}_c\right)\lessapprox
A_L\int^{D}_{0}\Xi_Lf_{r_L}(r)dr+A_N\int^{+\infty}_{0}\Xi_Nf_{r_N}(r)dr,
\end{align}
where
\begin{align}
&\Xi_L\triangleq\sum^{N_L}_{n=1}(-1)^{n+1}\binom{N_L}{n}\textrm{exp}\left(-\Theta_{L}(n)r^{\alpha_L}N_0\right)
\prod_{k=1}^3\varpi_{k}(M_s,\phi,n)\prod_{k=1}^3\varpi_{k}(M_a,1-\phi,n),
\\
&\Xi_N\triangleq\sum^{N_L}_{n=1}(-1)^{n+1}\binom{N_L}{n}\textrm{exp}\left(-\Theta_{N}(n)r^{\alpha_N}N_0\right)
\prod_{k=1}^3\varphi_{k}(M_s,\phi,n)\prod_{k=1}^3\varphi_{k}(M_a,1-\phi,n),
\\
&\varpi_1(a,b,n)\triangleq\Delta(C,r,D,\Theta_{L}(n),r^{\alpha_L},a,b),
\nonumber\\
&\varpi_2(a,b,n)\triangleq\Delta\left(1-C,{\min\left(\left(\frac{C_N}{C_L}\right)^{\frac{1}{\alpha_N}}r^{\frac{\alpha_L}{\alpha_N}},D\right)},D,\Theta_{L}(n),r^{\alpha_L},a,b\right)
,\nonumber\\
&\varpi_3(a,b,n)\triangleq\Delta\left(1,{\max\left(\left(\frac{C_N}{C_L}\right)^{\frac{1}{\alpha_N}}r^{\frac{\alpha_L}{\alpha_N}},D\right)},+\infty,\Theta_{L}(n),r^{\alpha_L},a,b\right)
,\nonumber\\
&\varphi_1(a,b,n)\triangleq\Delta\left(C,{\min\left(\left(\frac{C_L}{C_N}\right)^{\frac{1}{\alpha_L}}r^{\frac{\alpha_N}{\alpha_L}},D\right)},D,\Theta_{N}(n),r^{\alpha_N},a,b\right),
\nonumber\\
&\varphi_2(a,b,n)\triangleq\Delta\left(1-C,{\min\left(r,D\right)},D,\Theta_{N}(n),r^{\alpha_N},a,b\right),
\nonumber\\
&\varphi_3(a,b,n)\triangleq\Delta\left(1,\min(r,D),+\infty,\Theta_{N}(n),r^{\alpha_N},a,b\right),
%\nonumber\\
%&\Delta_{L,1}(n,a,b)\triangleq\textrm{exp}\left(-{2\pi C\lambda_Bp_{G_b(\theta)}(a)}\int^{D}_rF\left(\Theta_{L}(n),r^{\alpha_L},a,b\right)ydy\right)
%\nonumber\\
%&\Delta_{L,2}(n,a,b)\triangleq\textrm{exp}\left(-{2\pi (1-C)\lambda_Bp_{G_b(\theta)}(a)}\int^{D}_{\min\left(\left(\frac{C_N}{C_L}\right)^{\frac{1}{\alpha_N}}r^{\frac{\alpha_L}{\alpha_N}},D\right)}
%F\left(\Theta_{L}(n),r^{\alpha_L},a,b\right)
%ydy\right)
%\nonumber\\
%&\Delta_{L,3}(n,a,b)\triangleq\textrm{exp}\left(-{2\pi \lambda_Bp_{G_b(\theta)}(a)}\int^{+\infty}_{\max\left(\left(\frac{C_N}{C_L}\right)^{\frac{1}{\alpha_N}}r^{\frac{\alpha_L}{\alpha_N}},D\right)}
%F\left(\Theta_{L}(n),r^{\alpha_L},a,b\right)ydy\right)
%\\
%&\Delta_{N,1}(n,a,b)\triangleq\textrm{exp}\left(-{2\pi C\lambda_Bp_{G_b(\theta)}(a)}\int^{D}_{\min\left(\left(\frac{C_L}{C_N}\right)^{\frac{1}{\alpha_L}}r^{\frac{\alpha_N}{\alpha_L}},D\right)}F\left(\Theta_{N}(n),r^{\alpha_N},a,b\right)ydy\right)
%\\
%&\Delta_{N,2}(n,a,b)\triangleq\textrm{exp}\left(-{2\pi (1-C)\lambda_Bp_{G_b(\theta)}(a)}\int^{D}_{\min\left(r,D\right)}F\left(\Theta_{N}(n),r^{\alpha_N},a,b\right)ydy\right)
%\\
%&\Delta_{N,3}(n,a,b)\triangleq\textrm{exp}\left(-{2\pi \lambda_Bp_{G_b(\theta)}(a)}\int^{+\infty}_{\max\left(r,D\right)}F\left(\Theta_{N}(n),r^{\alpha_N},a,b\right)ydy\right)
\end{align}
$a_L$ and $a_N$ have been defined in Theorem 4, the analysis results of $f_{r_L}(r)$ and $f_{r_N}(r)$ have been given in Lemma 3,  $\Theta_{L}(n)\triangleq\frac{{T}_c n a_L}{\phi P_tM_s C_L}$, $\Theta_{N}(n)\triangleq\frac{{T}_c n a_N}{\phi P_tM_s C_N}$, $F(c,d,a,b) \triangleq1-\frac{1}{\left(1+cdab P_tC_Ly^{-\alpha_L}\right)^{N_L}}
$, $\Delta(z,v,y,c,d,a,b)\triangleq\textrm{exp}\left(-{2\pi z\lambda_B\textrm{Pr}_{x}(a)}\int^{y}_vF\left(c,d,a,b\right)ydy\right)$,
\end{Th}
\begin{IEEEproof}
The proof is given in Appendix F.
\end{IEEEproof}
\begin{figure}[!t]
\centering
\includegraphics[width=2.5in]{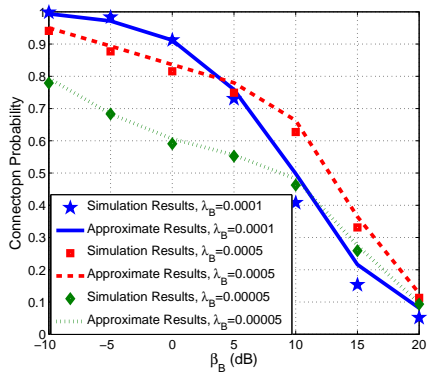}
\caption{Connection probability of mmWave communication with AN vs $T_c$. The system parameters are $P_t=30$ dB, $\phi=0.5,\theta_b=9, M_s=15$dB, $M_a=3$dB, $\beta_L=61.4$ dB, $\alpha_L=2,$ $\beta_N=72$ dB, $\alpha_N=2.92$, BW = 2 GHz, $\mathcal{F}_{dB}=10$, $C=0.12$, and $D=200$ m.}
\label{ConnectionprobabilityArtificialNoise}
\end{figure}
%Theoretical result in Theorem 6 is validated in Fig. \ref{ConnectionprobabilityArtificialNoise}.
In Fig. \ref{ConnectionprobabilityArtificialNoise}, we plot the connection probability $p_{con}$ versus $T_c$. From Fig. \ref{ConnectionprobabilityArtificialNoise}, we can find that approximate results coincide with  simulation ones well, which show that the approximate analysis result given in Theorem 6 is tight. In addition, from  simulation results, we can find an interesting phenomenon that for some $T_c$, a larger $\lambda_B$ may result in a smaller $p_{con}$. Therefore, we can conclude that $p_{con}$ is not a monotonically  increasing function of $\lambda_B$ for the whole range of $T_c$. This can be explained by the fact that although the distance from the authorized user to its serving BS decreases with the increasing $\lambda_B$, the network interference also increases. Therefore,  increasing $\lambda_B$ may not always improve $p_{con}$. This further shows that the  mmWave network with AN is interference-limited.
\subsection{Secrecy Probability}
In this subsection, we characterize the secrecy probability of the AN assisted mmWave communication. With the approximation (\ref{SectorAntennaModelArtificialNoise}), only  eavesdroppers inside the intended sector of the serving BS would wiretap the confidential information. Those eavesdroppers form a
fan-shaped PPP and by the mapping theorem \cite{wirelessnetworks}, they can be mapped as a homogeneous PPP on the whole
plane, denoted by $\Phi_Z$ with density ${\lambda_E}\textrm{Pr}_{x}(M_s)$.
%The SINR received by the eavesdropper at $z\in\Phi_Z$ is
%\begin{align}
%\textrm{SINR}_{E_z}=\frac{\phi P_tM_Sh_{X^*z}L\left(X^*,z\right)}{I_z+N_0},
%\end{align}
%where the interference from multiple interfering BSs $I_z=\sum_{y\in\Phi_I/{x^*}}M_S\phi P_th_{yx^*}L(y,x^*)+\sum_{y\in\Phi_A}M_a(1-\phi) P_th_{yx^*}L(y,x^*)$.
Since we consider the worst-case where each eavesdropper can eliminate the interference due to the information signals from other interfering BSs, only the AN would deteriorate the receiving performance of eavesdroppers.
%By the mapping theorem and the approximation (\ref{SectorAntennaModelArtificialNoise}), the BSs sending AN towards the eavesdropper at $z\in\Phi_Z$ form a homogeneous
%PPP $\Phi_{A_z}$ with intensity ${\lambda_Bp_{G_b(\theta)}(M_a)}$  on the whole
%plane.
Then, the received SINR by the eavesdropper at $z$ can be calculated as
\begin{align}
\textrm{SINR}_z=\frac{\phi P_tM_SL(x^*,z)g_{x^*z}}{I_{A_z}+N_0},\label{SINREZ}
\end{align}
where $I_{A_z}=\sum_{y\in\Phi_{A}}\left(1-\phi\right)P_tM_ag_{y,z}L(y,z)$.

In the case of non-colluding eavesdroppers, the secrecy probability of the mmWave network with AN can be calculated as $p_{sec}=\mathbb{E}_{\Phi_Z,\Phi_{A}}\left(\prod_{z\in\Phi_Z}\textrm{Pr}\left(\textrm{SINR}_z\leq T_e\right)\right)$, which is characterized by the following theorem.
\begin{Th}
The secrecy probability can be tightly lower bounded by
\begin{align}
p_{sec}\gtrapprox&\textrm{exp}\left(-2\pi C\lambda_E\textrm{Pr}_{x}(M_s)\int^{D}_0\Omega_L(r)rdr-2\pi (1-C)\lambda_E\textrm{Pr}_{x}(M_s)\int^{D}_0\Omega_N(r)rdr\right)
\nonumber\\
&\textrm{exp}\left(-2\pi \lambda_E\textrm{Pr}_{x}(M_s)\int^{+\infty}_D\Omega_N(r)rdr\right),
\end{align}
where $a_j\triangleq\left(N_j!\right)^{-\frac{1}{N_j}}$,
$
\Omega_j(r)\triangleq\sum^{N_j}_{n=1}(-1)^{n+1}\binom{N_j}{n}\textrm{exp}\left(-\frac{na_jT_er^{\alpha_j}N_0}{\phi P_tM_sC_j}\right)
\textrm{exp}\left(\Psi\left(\frac{(1-\phi)na_jT_er^{\alpha_j}}{\phi M_sC_j}\right)\right),\quad j=\{L,N\},
$
 and
\begin{align}
\Psi(s)\triangleq&-s\left(2\pi\lambda_B\sum_{j\in\{L,N\}}q_j\left(\textrm{Pr}_{x}(M_a)\frac{\left(M_aC_j\right)^{\frac{2}{\alpha_j}}}{\alpha_j}\sum^{N_j-1}_{m=0}\frac{\left(D^{\alpha_j}/(yC_j)\right)^{m+\frac{2}{\alpha_j}}}{\left(m+\frac{2}{\alpha_j}\right)\left(s+{D^{\alpha_j}/(yC_j)}\right)^{m+1}}
\right.\right.
\nonumber\\
&\left.\left.
{_2}F_1\left(1,m+1;m+\frac{2}{\alpha_j}+1;\frac{{D^{\alpha_j}/(M_aC_j)}}{{D^{\alpha_j}/(M_aC_j)}+s}\right)\right)+
2\pi\lambda_E\textrm{Pr}_{x}(M_a)\frac{\left(M_aC_N\right)^{\frac{2}{\alpha_N}}}{\alpha_N}
\right.\nonumber\\
&\left.\sum^{N_N-1}_{m=0}
\frac{\left(D^{\alpha_N}/(M_aC_N)\right)^{m+\frac{2}{\alpha_N}}}{\left(1-\frac{2}{\alpha_N}\right)\left(s+D^{\alpha_N}/(M_aC_N)\right)^{m+1}}
{_2}F_1\left(1,m+1;2-\frac{2}{\alpha_N};\frac{s}{s+D^{\alpha_N}/(M_aC_N)}\right)
\right).
\end{align}
\end{Th}
\begin{IEEEproof}
The proof is given in Appendix G.
\end{IEEEproof}

\begin{figure}[!t]
\centering
\includegraphics[width=2.5in]{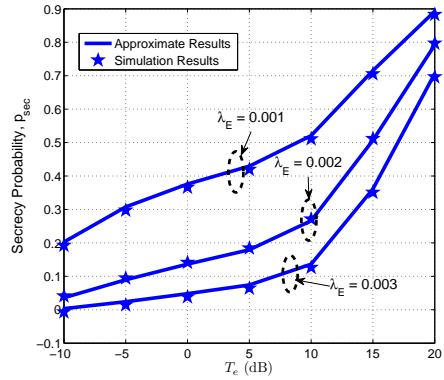}
\caption{Secrecy probability of mmWave communication with AN versus $T_e$. The system parameters are $P_t=30$ dB, $\theta_b=9^o$, $M_s=15$dB, $M_a=3$dB, BW = 2 GHz, $\beta_L=61.4$ dB, $\alpha_L=2,$ $\beta_N=72$ dB, $\alpha_N=2.92$, $\mathcal{F}_{dB}=10$, $\phi=0.5$, $P_t=30$dB, $\lambda_B=0.00005,C=0.12$, and $D=200$ m.}
\label{SecrecyProbabilityValidation}
\end{figure}
Theoretical results in Theorem 7 are validated in Fig. \ref{SecrecyProbabilityValidation}. In Fig. \ref{SecrecyProbabilityValidation}, we plot the secrecy probability $p_{sec}$ versus $T_e$. From Fig. \ref{SecrecyProbabilityValidation}, we can find that the approximate results coincide with the simulation ones well, which show that the lower bound given in Theorem 7 is tight.

\section{Simulation Result}
In this section, more representative simulation results are
provided to characterize the secrecy performance of mmWave networks and the effect of different network parameters. Considering mmWave networks operating at a carrier frequency $F_c=28$ GHz, the path-loss model are taken from \cite[Tables I]{MillimeterChannelModeling}. Specially, the transmission bandwidth BW = 2 GHz, the noise figure $\mathcal{F}_{dB}=10$, the BS's transmit power $P_t=30$ dB, the Nakagami fading parameters of the LOS (NLOS) link are $N_L=3$ ($N_N=2$), and the path-loss model: $\beta_L=61.4$ dB, $\alpha_L=2,$ $\beta_N=72$ dB,  $\alpha_N=2.92$.
Since the theoretical analysis results obtained in this paper have been validated by the simulation results in Fig. \ref{noncolludingEvessecureconnectivity}-Fig. \ref{SecrecyProbabilityValidation}, all of the simulation results in this section are theoretical analysis results.

\subsection{Secrecy performance evaluation of noise-limited mmWave cellular networks}
In this subsection, employing analysis results in Section III, we  illustrate the secrecy performance of noise-limited mmWave networks in the presence of \textbf{non-colluding} and \textbf{colluding} eavesdroppers.
\begin{figure}[!t]
\centering
\includegraphics[width=2.5in]{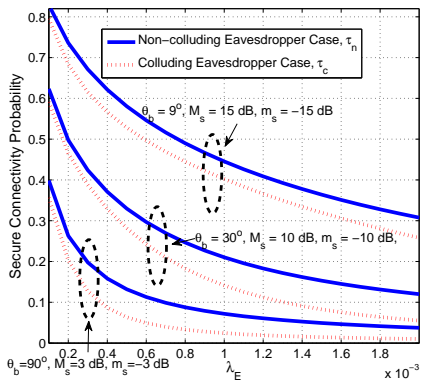}
\caption{Secrecy connectivity probability of the noise-limited mmWave communication in the presence of multiple eavesdroppers versus $\lambda_E$. The system parameters are  $\lambda_B=0.00005,C=0.081$, and $D=250$ m.}
\label{SecrecyconnectivityVersusLambdaE}
%\centering
%\includegraphics[width=2.5in]{SCPversusLambdaB}
%\caption{Secrecy connectivity probability  of the noise-limited mmWave communication in the presence of multiple eavesdroppers versus $\lambda_B$. The system parameters are
%%BW = 2 GHz, $P_t=30$ dB, $\mathcal{F}_{dB}=10$, $\beta_L=61.4$ dB, $\alpha_L=2,$ $\beta_N=72$ dB, $\alpha_N=2.92$,
%$\lambda_E=0.0005,C=0.081$, and $D=250$ m.}
%\label{SecrecyconnectivityVersusLambdaB}
\end{figure}

Fig. \ref{SecrecyconnectivityVersusLambdaE} plots the secrecy connectivity probability of the  mmWave communication in the presence of multiple non-colluding and colluding eavesdroppers versus $\lambda_E$. Obviously, the wiretapping capability of the colluding eavesdroppers is larger than the non-colluding case. Therefore, compared with the non-colluding eavesdroppers, the secrecy connectivity probability for the colluding case deteriorates. With the increasing $\lambda_E$, the wiretapping capability of eavesdroppers increases and the secrecy connectivity probability decreases. Furthermore, the secrecy performance would be improved with the improving directionality of the beamforming of each BSs. This can be explained by the fact that the high gain narrow beam antenna decreases the information leakage, improves the receive performance of the authorized user, and increases the secure connectivity probability.

%Fig. \ref{SecrecyconnectivityVersusLambdaB} plots the secrecy connectivity probability  versus $\lambda_B$. Secrecy connectivity probability improves with the increasing $\lambda_B$. Since with the increasing $\lambda_B$, the link length from the serving BS to authorized users decreases, and for the noise-limited scenario, without considering the network interference, the receive performance of authorized users would be improved with the increasing $\lambda_B$ and the secrecy connectivity probability increases.

\begin{figure}[!t]
\centering
\includegraphics[width=2.5in]{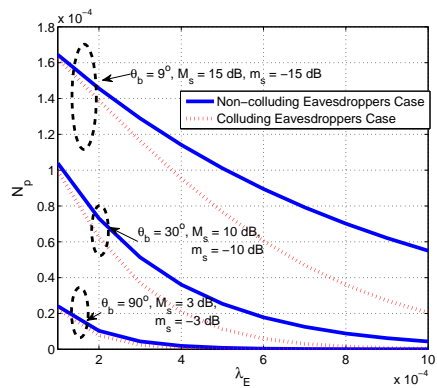}
\caption{Average number of perfect communication links per unit, $N_p$ of the noise-limited mmWave communication in the presence of multiple eavesdroppers versus $\lambda_E$. The system parameters are $P_t=30$ dB, $T_c$=10dB, $T_e$=0 dB,
%BW = 2GHz, $\mathcal{F}_{dB}=10$, $\beta_L=61.4$ dB, $\alpha_L=2,$ and $\beta_N=72$ dB, $\alpha_N=2.92$,
$\lambda_B=0.0002,C=0.12$, and $D=200$ m.}
\label{NumberNPversusLambdaE}
\end{figure}

Fig. \ref{NumberNPversusLambdaE} plots  $N_p$  versus $\lambda_E$. Compared with the non-colluding eavesdroppers, the performance deterioration of the colluding case increases with the increasing $\lambda_E$, especially for the BS equipped with highly directional antenna arrays. Furthermore, the simulation results show that the directional beamforming is very important for the secrecy communication. For example, for the non-colluding eavesdroppers case, when $\theta_b=9^o,M_s=15$dB, $M_a=3$dB, $\lambda_E=4\times 10^{-4}$, $N_p\approx 1.1\times 10^{-4}$ and more than half of communication links is perfect, on average. However, for other two cases of array patterns, $N_p$ reduces greatly due to the increasing beam width of the main lobe and the decreasing array gains of the intended sector.

The simulation results above show that the directional beamforming of BSs is very important for the secrecy performance of noise-limited mmWave networks. Therefore, in practice, for improving the security of noise-limited mmWave networks, BSs should perform the highly directional beamforming.

\subsection{Secrecy performance evaluation of interference-limited mmWave cellular networks with AN}
In this subsection, employing analysis results in Section IV, we illustrate the impact of the AN on the secrecy performance of interference-limited mmWave networks in the presence of  \textbf{non-colluding} eavesdroppers.

\begin{figure}[!t]
\centering
\includegraphics[width=2.5in]{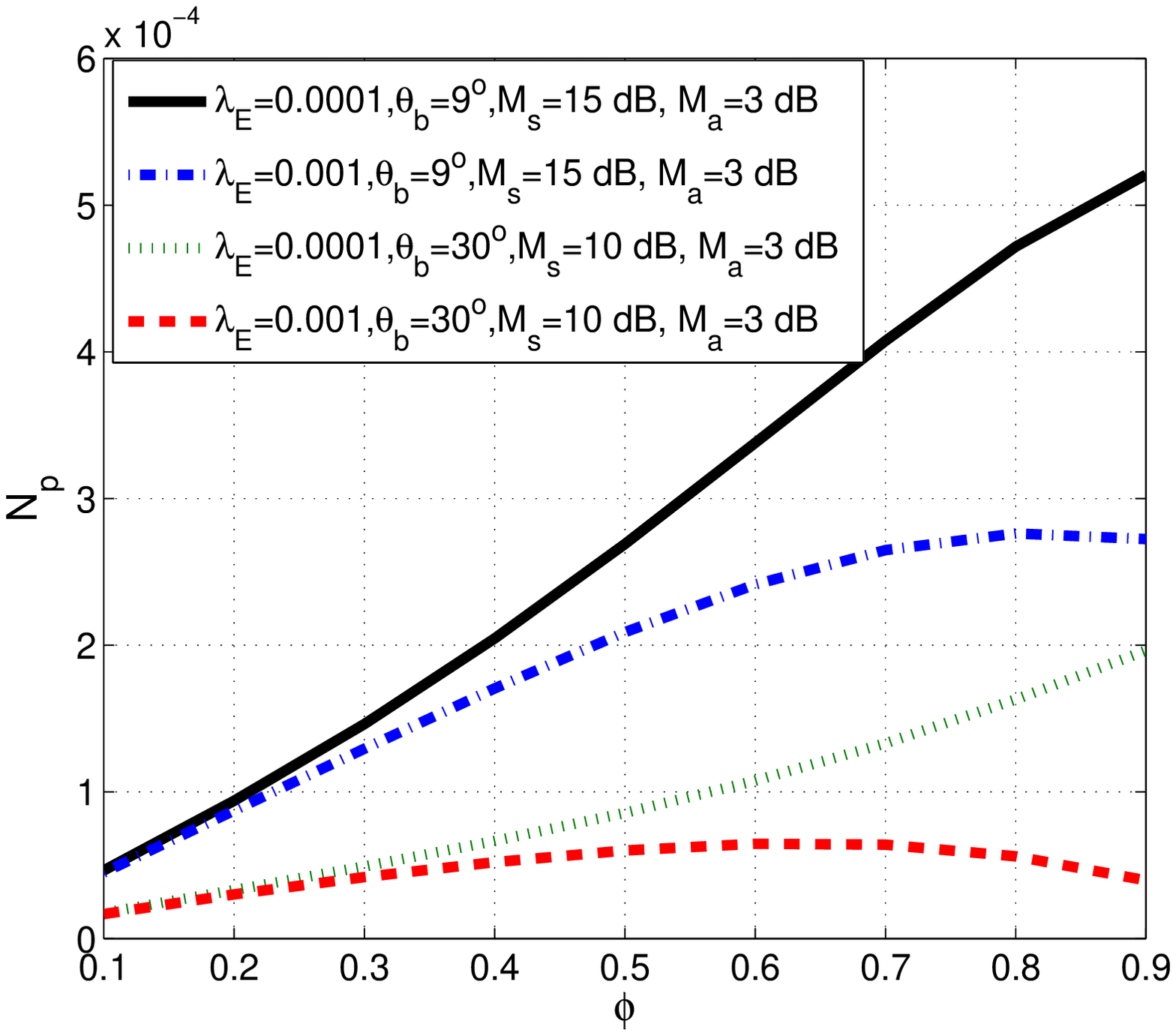}
\caption{Average number of perfect communication links per unit, $N_p$  of the interference-limited mmWave communication in the presence of multiple non-colluding eavesdroppers versus the power allocation coefficient $\phi$. The system parameters are $T_c$=10dB, $T_e$=0 dB,
%BW = 2 GHz, $P_t=30$ dB, $\mathcal{F}_{dB}=10$, $\beta_L=61.4$ dB, $\alpha_L=2,$ and $\beta_N=72$ dB, $\alpha_N=2.92$,
$\lambda_B=0.0008,C=0.12$, and $D=200$ m.}
\label{NumberNPArtificialNoiseVersusPhi}
%\end{figure}
%\begin{figure}[!t]
%\centering
\includegraphics[width=2.5in]{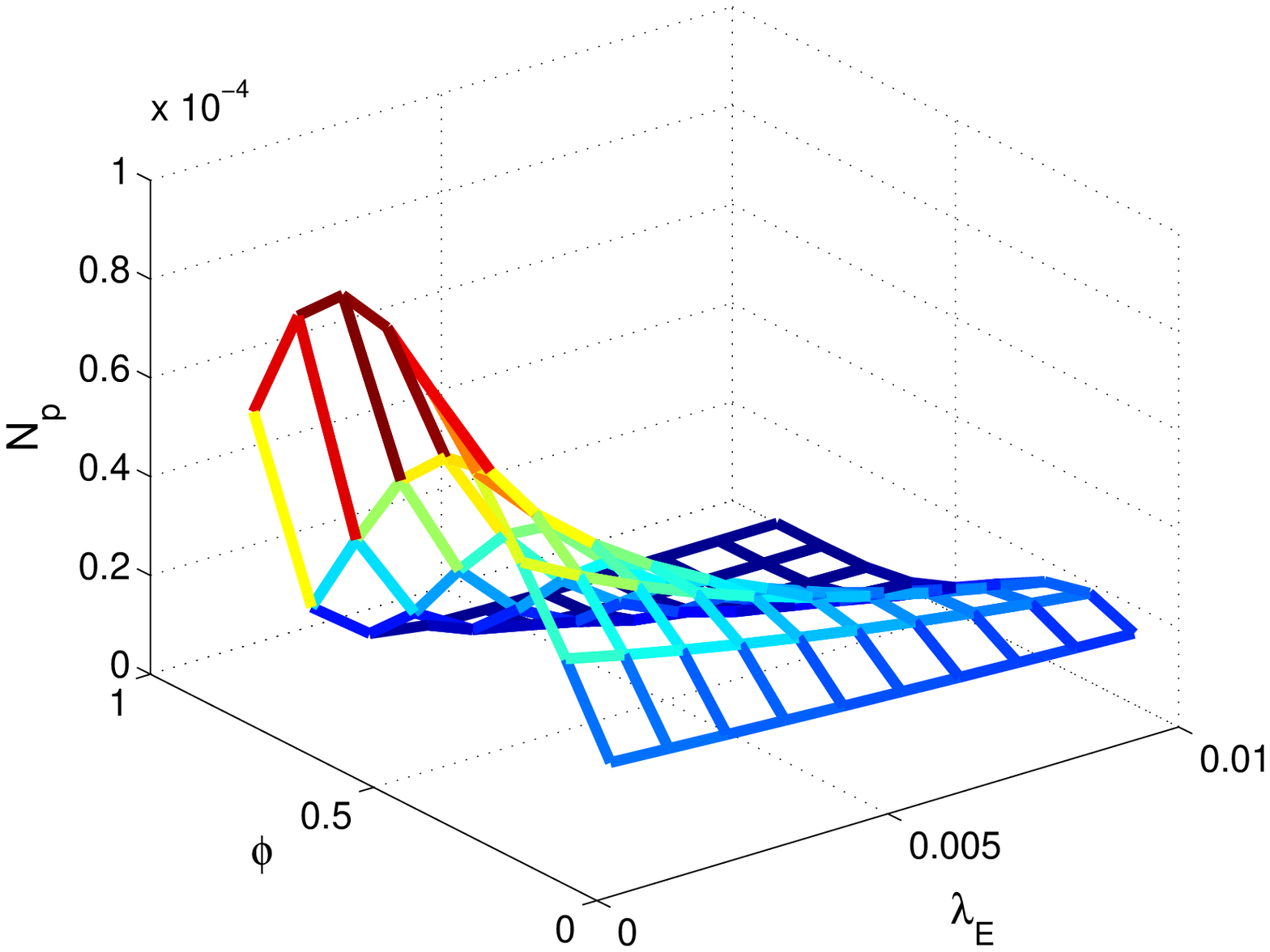}
\caption{Average number of perfect communication links per unit, $N_p$  of the interference-limited mmWave communication in the presence of multiple non-colluding eavesdroppers versus the power allocation coefficient $\phi$ and $\lambda_E$. The system parameters are $\theta_b=30^o$, $M_s=10$ dB, $M_a=3$ dB, $T_c$ = 10dB,  $T_e$ = 0 dB,
%BW = 2 GHz, $P_t=30$ dB, $\mathcal{F}_{dB}=10$, $\beta_L=61.4$ dB, $\alpha_L=2,$ and $\beta_N=72$ dB, $\alpha_N=2.92$,
$\lambda_B=0.001,C=0.12$, and $D=200$ m.}
\label{TwoDimensionalNp}
\end{figure}

\begin{figure}[!t]
\centering
\includegraphics[width=2.5in]{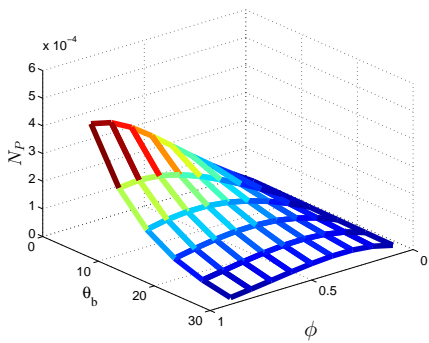}
\caption{Average number of perfect communication links per unit, $N_p$  of the interference-limited mmWave communication in the presence of multiple non-colluding eavesdroppers versus the power allocation coefficient $\phi$ and $\theta_b$. The system parameters are $P_t=30$ dB, $M_s=15$ dB, $M_a=3$ dB, $T_c$=10dB, $T_e$=0 dB,
$\lambda_B=0.0008$, $\lambda_E=0.001$, $C=0.12$, and $D=200$ m.}
\label{NumberNPversusThetab}
\end{figure}

Fig. \ref{NumberNPArtificialNoiseVersusPhi} plots $N_p$ achieved by the mmWave communication  versus the power allocation coefficient $\phi$ for different antenna patterns and $\lambda_E$. From the simulation results in Fig. \ref{NumberNPArtificialNoiseVersusPhi}, we find that the optimal fraction of the power allocated to the AN decreases with the decreasing $\lambda_E$ and improving directivity of the antenna array equipped at each BS. This can be explained by the fact  that with the decreasing $\lambda_E$, the wiretapping capability of eavesdroppers decreases, and the optimal fraction of the power allocated to the AN can be reduced. Accordingly, with the highly directional beamforming, the information leakage decreases, and the receiving performance of eavesdroppers decreases. Therefore, the power allocated to the AN can be reduced.
For showing the effect of $\lambda_E$ and $\theta_b$ on the optimal $\phi$  further, we plot $N_p$ versus $\phi$ and $\lambda_E$ in Fig. \ref{TwoDimensionalNp}, and $N_p$ versus $\phi$ and $\theta_b$ in Fig. \ref{NumberNPversusThetab}. From the simulation results, it is clear that the optimal $\phi$ for maximizing $N_P$ increases with the decreasing $\lambda_E$ and decreasing $\theta_b$, which validates the conclusions draw above.

Simulation results show that the optimal power allocated to AN depends on $\lambda_E$ and antenna pattern. The highly directional antenna array and small $\lambda_E$ both would decrease the power allocated to AN.

\subsection{Secrecy performance comparison between the AN assisted microwave network and the AN assisted mmWave network}

\begin{figure}[!t]
\centering
\includegraphics[width=2.5in]{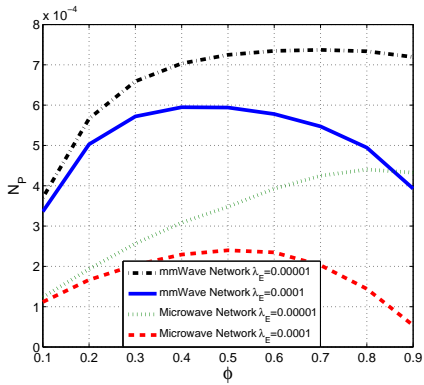}
\caption{Secrecy performance comparison between the microwave network and mmWave network. The system parameters of the mmWave network are $P_t=30$ dB, $T_c$=0dB, $T_e$=-30 dB, $\theta_b=9^o$, $M_s=15$ dB, $M_a=3$ dB, $\lambda_B=0.0008,C=0.12$, and $D=200$ m. The system parameters of the microwave network are $P_t=30$ dB, $T_c$=0dB, $T_e$=-30 dB, the beam width of the confidential signal is $60^o$, the beam width of AN is $300^o$, the number of antennas equipped at each microwave BS is 6, $\lambda_B=0.0008$, the small scale channel fading follows the normalized Rayleigh fading.
%BW = 2GHz, $\mathcal{F}_{dB}=10$, $\beta_L=61.4$ dB, $\alpha_L=2,$ and $\beta_N=72$ dB, $\alpha_N=2.92$,
.}
\label{ComparingwithMicrowaveNetwork}
\end{figure}

For validating the secrecy performance of the mmWave communication, we perform the secrecy performance comparison between the microwave network  and mmWave network in Fig. \ref{ComparingwithMicrowaveNetwork}, where the intensities of microwave BSs and  mmWave BSs are both set to be 0.0008.
The carrier frequency of the microwave communication is $F_c=2.5$ GHz.
Just as \cite{Enhancing,MicrowaveCommunication}, the antenna pattern of the microwave BS is approximated by (\ref{SectorAntennaModelArtificialNoise}), where the beam width of the confidential signal is set to be $60^o$, and the beam width of AN is set to be $300^o$, the small scale channel fading of the microwave communication is the normalized Rayleigh fading. The large-scale path loss of the urban area cellular radio communication is used to model the path-loss of the microwave communication \cite{3GPP}, where the  path-loss exponent is 2.7.
Since the microwave network is interference-limited \cite{Enhancing}, the received noise power is ignored in the simulation. Therefore, only the ratio between the antenna gain of the confidential signals and the antenna gain of AN would determine the SINR received at the typical authorized user and eavesdropper in the microwave network \cite{Enhancing}. Just as \cite{Enhancing}, the ratio between the antenna gain of the confidential signals and the antenna gain of AN in the microwave BS is set to be $M-1$, where $M$ is the number of antennas equipped at the microwave BS. In the simulation results of Fig. \ref{ComparingwithMicrowaveNetwork}, we set $M=6$.
From the simulation results in Fig. \ref{ComparingwithMicrowaveNetwork}, we can find that the mmWave network can achieve better secrecy performance than the microwave network. This is because the unique characteristic of the mmWave communication: blockage effects and highly directional beamforming antenna arrays. Due to blockage effects, the wiretapping capability of eavesdroppers would decreases, since the blockage effects would deteriorate the reception quality of a large portion of  eavesdroppers. Due to the highly directional beamforming antenna arrays equipped at each mmWave BS, the reception quality of the intended receiver would be improved, and the confidential information leakage would be decreased.

\section{Conclusions}
In this paper, considering distinguishing features of the mmWave cellular network, we characterize the secrecy performance of the noise-limited mmWave network and the AN-assisted mmWave network.
For the noise-limited case, we analyze the secure connectivity probability and average number of perfect communication links per unit area for colluding and non-colluding eavesdroppers. For the AN-assisted mmWave network which is interference-limited,  by taking the network interference into consideration, we characterize the distributions of the received SINRs at the intended receiver and eavesdroppers, and  average number of perfect communication links per unit area for non-colluding eavesdroppers. Simulation results show that the array pattern and intensity of eavesdroppers are very important system parameters for improving the secrecy performance of the mmWave communication. In particular, for the AN-assisted mmWave networks, the power allocated to AN depends on the array pattern and the intensity of eavesdroppers. It decreases with the decreasing beam width of the main lobe and decreasing intensity of eavesdroppers.

\appendices
\section{Proof of Lemma 1}
We first show the derivation of $f_{d_ L^*}(r)$.
Given the typical authorized user observes at least one LOS
BS, the complementary cumulative distribution function (CCDF) of $d_ L^*$ can be derived as
\begin{align}
\textrm{Pr}\left(d_ L^*\geq r\right)&\triangleq\textrm{Pr}\left(\Phi_{B_L}\left(B(o,r)\right)=0|\Phi_{B_L}\left(B(o,D)\right)\neq 0\right)
=
\frac{\textrm{e}^{-C\lambda_B\pi r^2}\left(1-\textrm{e}^{-C\lambda_B\pi \left(D^2-r^2\right)}\right)}{1-\textrm{e}^{-C\lambda_B\pi D^2}},\label{CDFDL}
\end{align}
Then, with (\ref{CDFDL}), the pdf $f_{d_ L^*}(r)=-\frac{d \textrm{Pr}\left(d_ L^*\geq r\right)}{dr}$ that can be derived as (\ref{pdfdl}).

%In the following, we first derive the CCDF $\textrm{Pr}\left(d^*_N\geq r\right)$ and the pdf $f_{d_ N^*}(r)=-\frac{d \textrm{Pr}\left(d^*_N\geq r\right)}{dr}$.
Secondly, invoking the PPP's void probability \cite{wirelessnetworks}, the CCDF $\textrm{Pr}\left(d^*_N\geq r\right)$ can be derived as
\begin{align}
&\textrm{Pr}\left(d^*_N\geq r\right)=
\nonumber\\
&\textrm{Pr}\left(\Phi_{B_N}\left(B(o,r)\right)=0\right)\mathbb{I}(r\leq D)+\textrm{Pr}\left(\Phi_{B_N}\left(B(o,D)\right)=0,\Phi_{B_N}\left(B(o,r)/B(o,D)\right)=0\right)
\mathbb{I}(r> D)
\nonumber\\
&=\textrm{exp}\left(\left(1-C\right)\lambda_B\pi r^2\right)\mathbb{I}(r\leq D)
+\textrm{exp}\left(\left(1-C\right)\lambda_B\pi D^2\right)
\textrm{exp}\left(-\lambda_B\pi \left(r^2-D^2\right)\right)
\mathbb{I}(r> D).
\end{align}
Finally, calculating $-\frac{d \textrm{Pr}\left(d^*_N\geq r\right)}{dr}$, the pdf $f_{d_ N^*}(r)$ can be derived as (\ref{distancedistribution}).

\section{Proof of Lemma 2}
We do the following derivations.
\begin{align}
A_N&=\textrm{Pr}\left(C_L(d^*_L)^{-\alpha_L}\leq C_N(d^*_N)^{-\alpha_N}\right)\textrm{Pr}\left(\Phi_{B_L}(B(o,D))\neq 0\right)+\textrm{Pr}\left(\Phi_{B_L}(B(o,D))= 0\right)
\nonumber\\
&={\textrm{Pr}\left(\left(\frac{C_L}{C_N}\right)^{\frac{1}{\alpha_L}}(d^*_N)^{\frac{\alpha_N}{\alpha_L}}\leq d^*_L\right)\left(1-\textrm{e}^{-\pi C\lambda_BD^2}\right)}+\textrm{e}^{-\pi C\lambda_BD^2}
\nonumber\\
&=
\mathbb{E}_{d^*_N\leq \left(\frac{C_L}{C_N}\right)^{-\frac{1}{\alpha_N}}D^{\frac{\alpha_L}{\alpha_N}}}\left({ \textrm{e}^{-\pi C\lambda_B\left(\frac{C_L}{C_N}\right)^{\frac{2}{\alpha_L}}(d_N^*)^{\frac{2\alpha_N}{\alpha_L}}}-\textrm{e}^{-\pi C\lambda_B D^2} }\right)
+{\textrm{e}^{-\pi C\lambda_B D^2}}.\label{ALproof}
\end{align}
Finally, (\ref{AN}) can be obtained by employing the pdf $f_{d^*_N}(r)$ in (\ref{distancedistribution}).
Accordingly, $A_L$ can be derived as
\begin{align}
&A_L=\textrm{Pr}\left(C_L(d^*_L)^{-\alpha_L}\geq C_N(d^*_N)^{-\alpha_N}\right)\textrm{Pr}\left(\Phi_{B_L}(B(o,D))\neq 0\right)
\nonumber\\
&=\left(\textrm{Pr}\left(D\geq\left(\frac{C_L}{C_N}\right)^{\frac{1}{\alpha_L}}\left(d^*_N\right)^{\frac{\alpha_N}{\alpha_L}}\geq d_L^*\right)+\textrm{Pr}\left(D\leq\left(\frac{C_L}{C_N}\right)^{\frac{1}{\alpha_L}}\left(d^*_N\right)^{\frac{\alpha_N}{\alpha_L}}\right)\right)\textrm{Pr}\left(\Phi_{B_L}(B(o,D))\neq 0\right)
\nonumber\\
&{=}\left(\int^\mu_0\left(\textrm{Pr}\left(\left(\frac{C_L}{C_N}\right)^{\frac{1}{\alpha_L}}r^{\frac{\alpha_N}{\alpha_L}}\geq d^*_L\right)\right)f_{d_N^*}(r)dr+\textrm{Pr}\left(d^*_N\geq\mu\right)\right)\textrm{Pr}\left(\Phi_{B_L}(B(o,D))\neq 0\right).\label{additonalproof}
\end{align}
Since $\textrm{Pr}\left(d_ L^*\leq r\right)=1-\textrm{Pr}\left(d_ L^*\geq r\right)$ and $\textrm{Pr}\left(d_ L^*\geq r\right)$ has been defined in (\ref{CDFDL}), substituting the analytical expression of $\textrm{Pr}\left(d_ L^*\leq r\right)$  and $f_{d_N^*}(r)$ in (\ref{distancedistribution}) into (\ref{additonalproof}), we obtain $A_L=1-A_N$.
\section{Proof of Lemma 3}
We first show the derivation of $f_{r_L}(r)$. The CCDF  $\textrm{Pr}\left(r_L\geq r\right)$ can be derived as
\begin{align}
&\textrm{Pr}\left(r_L\geq r\right)\triangleq\textrm{Pr}\left(\Phi_{B_L}(B(o,r))=0\left|C_L(d^*_L)^{-\alpha_L}\geq C_N(d^*_N)^{-\alpha_N} \right.\right)
\nonumber\\
&=
\frac{\textrm{Pr}\left(\Phi_{B_L}(B(o,D))\neq 0\right)\textrm{Pr}\left(r\leq d^*_L,C_L(d^*_L)^{-\alpha_L}\geq C_N(d^*_N)^{-\alpha_N}\right)}{\textrm{Pr}\left(C_L(d^*_L)^{-\alpha_L}\geq C_N(d^*_N)^{-\alpha_N}\right)}
\nonumber\\
&\overset{(a)}{=}\frac{1}{A_L}\int^{D}_r\textrm{Pr}\left(\Phi_N\cap b\left(o,\left(\frac{C_N}{C_L}\right)^{\frac{1}{\alpha_N}}y^{\frac{\alpha_L}{\alpha_N}}\right)=\emptyset\right)2\pi C\lambda_By\textrm{exp}\left(-\pi C\lambda_By^2\right)dy.\label{distancefrlproof}
\end{align}
Step $(a)$ can be derived by the PPP's void probability and $f_{d^*_L}(r)$.
Then calculating  $f_{r_L}(r)=-\frac{d \textrm{Pr}\left(r_L\geq r\right)}{dr}$, the pdf $f_{r_L}(r)$ can be derived as (\ref{frL}).

We show the derivation of $f_{r_N}(r)$. The CCDF $\textrm{Pr}\left(r_N\geq r\right)$ is equivalent to the conditional CCDF
\begin{align}
&\textrm{Pr}\left(r_N\geq r\right)\triangleq\textrm{Pr}\left(d^*_N\geq r\left| C_L(d^*_L)^{-\alpha_L}\leq C_N(d^*_N)^{-\alpha_N}\right.\right)=
\nonumber\\
&\frac{\textrm{Pr}\left(r\leq d^*_N,C_L(d^*_L)^{-\alpha_L}\leq C_N(d^*_N)^{-\alpha_N}\right)\textrm{Pr}\left(\Phi_{B_L}\left(B(o,D)\right)\neq 0\right)}{A_N}+\frac{\textrm{Pr}\left(r\leq d^*_N,\Phi_{B_L}\left(B(o,D)\right)= 0\right)}{A_N}.
\label{CCDFdN}
\end{align}
We first derive the first term in (\ref{CCDFdN}) as
\begin{align}
&\frac{\mathbb{E}_{d^*_N\geq r}\left(\textrm{Pr}\left(\left(\frac{C_L}{C_N}\right)^{\frac{1}{\alpha_L}}(d_N^*)^{\frac{\alpha_N}{\alpha_L}}\leq d^*_L\right)\right)\textrm{Pr}\left(\Phi_{B_L}\left(B(o,D)\right)\neq 0\right)}{A_N}
\label{Derivation1}\\
&\overset{(b)}{=}\frac{\mathbb{E}_{d^*_N\geq r}\left(\textrm{exp}\left(-C\lambda_B\pi\left(\frac{C_L}{C_N}\right)^{\frac{2}{\alpha_L}}(d_N^*)^{\frac{2\alpha_N}{\alpha_L}}\right)
-\textrm{exp}\left(-C\lambda_B\pi D^2\right)
\right)}{A_N}\mathbb{I}\left(r\leq \left(\frac{C_N}{C_L}\right)^{\frac{1}{\alpha_N}}D^{\frac{\alpha_L}{\alpha_N}}\right)
%\overset{(c)}{=}\\
%&\frac{\int^{\left(\frac{C_N}{C_L}\right)^{\frac{1}{\alpha_N}}D^{\frac{\alpha_L}{\alpha_N}}}_r\left(e^{-C\lambda_B\pi\left(\frac{C_L}{C_N}\right)^{\frac{2}{\alpha_L}}(y)^{\frac{2\alpha_N}{\alpha_L}}}
%-e^{-C\lambda_B\pi D^2}\right)2\pi (1-C)\lambda_Bye^{-\pi (1-C)\lambda_By^2}dy
%\mathbb{I}\left(r\leq \left(\frac{C_N}{C_L}\right)^{\frac{1}{\alpha_N}}D^{\frac{\alpha_L}{\alpha_N}}\right)}{A_N},
\label{DerivationdN}
\end{align}
%since the probability is nonzero if and only if $\left(\frac{C_L}{C_N}\right)^{\frac{1}{\alpha_L}}r^{\frac{\alpha_N}{\alpha_L}}\leq D$
Step $(b)$ is obtained according to the PPP's void probability.
%Step $(c)$ is obtained by averaging over $d_N^*$.

Finally, we derive the second term in (\ref{CCDFdN}). Since the LOS BS process and NLOS BS process are two independent PPPs, we have
\begin{align}
\frac{\textrm{Pr}\left(r\leq d^*_N,\Phi_{B_L}\left(B(o,D)\right)= 0\right)}{A_N}&=\frac{\textrm{exp}\left({-(1-C)\lambda_B\pi r^2-C\lambda_B\pi D^2}\right)}{{A_N}}\mathbb{I}\left(r\leq D\right)
+\frac{\textrm{e}^{-\lambda_B\pi r^2}}{A_N}\mathbb{I}\left(r\geq D\right).
\label{secondTerm}
\end{align}
Substituting (\ref{secondTerm}) and (\ref{DerivationdN}) into (\ref{CCDFdN}),  $f_{r_N}(r)=-\frac{d\textrm{Pr}\left(r_N\geq r\right)}{d r}$, which can be derived as (\ref{frN}).
\section{Proof of Lemma 4}
 The point process $\mathcal{N}_E$ can be regarded as a transformation of the point process $\Phi_E$ by the probability kernel
$
 p\left(z,A\right)=\textrm{Pr}\left(\frac{1}{G_b(\theta)g_{xz}L(x,z)}\in A\right),z\in\mathbb{R}^2,A\in\mathcal{B}(\mathbb{R}^+).
$
 According to the displacement theorem \cite{wirelessnetworks},  $\mathcal{N}_E$ is a PPP on $\mathbb{R}^+$ with the intensity measure $\Lambda_E\left(0,t\right)$ given by
\begin{align}
\Lambda_E\left(0,t\right)=\lambda_E\int_{\mathbb{R}^2}\textrm{Pr}\left(\frac{1}{G_b(\theta)g_{xz}L(x,z)}\in\left[0,t\right]\right)dz.
\label{lambdaE}
\end{align}

From the blockage model in Section II-B, we know that $\Phi_E$ is divided into two independent point processes, i.e., the LOS and NLOS eavesdropper processes. %In particular, when $\left|\left|z-x\right|\right|_F\leq D$, the intensity of the LOS eavesdroppers
%is $C\lambda_E$, and the intensity  of the NLOS eavesdroppers
%is $(1-C)\lambda_E$. When $\left|\left|z-x\right|\right|_F\geq D$, the intensity  of the LOS eavesdroppers process is 0, and the intensity  of the NLOS eavesdroppers
%is $\lambda_E$.
Furthermore, the directivity gains received at the eavesdroppers in the main and sidelobes are different. Therefore, considering these and changing to a polar
coordinate system, $\Lambda_E\left(0,t\right)$ in (\ref{lambdaE}) can be further derived as
\begin{align}
\Lambda_E\left(0,t\right)=&2\pi\lambda_E\sum_{V\in\{M_s,m_s\}}\textrm{Pr}_{G_b}(V)\sum_{j\in\{L,N\}}q_j\int^D_0\textrm{Pr}\left(\frac{r^{\alpha_j}}{G_b(\theta)g_rC_j}\leq t|G_b(\theta)=V\right)rdr
\nonumber\\
&+2\pi\lambda_E\sum_{V\in\{M_s,m_s\}}\textrm{Pr}_{G_b}(V)\int^{+\infty}_D\textrm{Pr}\left(\frac{r^{\alpha_N}}{G_b(\theta)g_rC_N}\leq t|G_b(\theta)=V\right)rdr,\label{DerivationofLambdaE}
\end{align}
where $g_r$ denotes the small-scale fading of eavesdropper which is $r$ distant from the target BS at $x$. $g_r\sim\textrm{gamma}\left(N_L,1\right)$ if the link between the eavesdropper and the target BS is LOS, otherwise, $g_r\sim\textrm{gamma}\left(N_N,1\right)$.

For getting the analysis result $\Lambda_E\left(0,t\right)$, the analysis results of integral formulaes in (\ref{DerivationofLambdaE})  should be derived. Firstly, the integral $\int^D_0\textrm{Pr}\left(\frac{r^{\alpha_j}}{G_b(\theta)g_rC_j}\leq t|G_b(\theta)=V\right)rdr$ can be derived with the procedures in (\ref{analysisIntegral1}).
\begin{align}
&\int^D_0\textrm{Pr}\left(g_r\geq \frac{r^{\alpha_j}}{VC_jt}\right)rdr\overset{(a)}{=}
\int^D_0\left(1-\frac{\gamma\left(N_j,\frac{r^{\alpha_j}}{VC_jt}\right)}{\Gamma(N_j)}\right)rdr
\nonumber\\
&\overset{(b)}{=}\int^D_0\frac{\Gamma\left(N_j,\frac{r^{\alpha_j}}{VC_jt}\right)}{\Gamma(N_j)}rdr\overset{(c)}{=}\int^D_0e^{-\frac{r^{\alpha_j}}{VC_jt}}\sum^{N_j-1}_{m=0}\left(\frac{r^{\alpha_j}}{VC_jt}\right)^m\frac{1}{m!}rdr
\overset{(d)}{=}\frac{\left(VC_jt\right)^{\frac{2}{\alpha_j}}}{\alpha_j}\sum^{N_j-1}_{m=0}\frac{\gamma\left(m+\frac{2}{\alpha_j},\frac{D^{\alpha_j}}{VC_jt}\right)}{m!}.
\label{analysisIntegral1}
\end{align}
Step (a) is due to $g_r\sim\textrm{gamma}\left(N_j,1\right)$, step (b) is due to \cite[eq.(8.356.3)]{Table}, step (c) is due to \cite[eq.(8.352.2)]{Table}, and step (d) is due to \cite[eq.(3.381.1)]{Table}.

With a similar procedure,  the integral $\int^{+\infty}_D\textrm{Pr}\left(\frac{r^{\alpha_j}}{G_b(\theta)g_rC_j}\leq t|G_b(\theta)=V\right)rdr$ can be derived as
\begin{align}
\int^{+\infty}_D\textrm{Pr}\left(\frac{r^{\alpha_j}}{G_b(\theta)g_rC_j}\leq t|G_b(\theta)=V\right)rdr=\frac{\left(VC_jt\right)^{\frac{2}{\alpha_j}}}{\alpha_j}\sum^{N_j-1}_{m=0}\frac{\Gamma\left(m+\frac{2}{\alpha_j},\frac{D^{\alpha_j}}{VC_jt}\right)}{m!}.
\label{analysisIntegral2}
\end{align}
Finally, substituting (\ref{analysisIntegral1}) and (\ref{analysisIntegral2}) into (\ref{DerivationofLambdaE}), the proof can be completed.
\section{Proof of Theorem 3}
The achievable secure connectivity probability, $\tau_{c}$ can be calculated as
\begin{align}
&\tau_{c}\overset{(g)}{=}
\mathbb{E}_{r_j}\left[\mathbb{E}_{I_E}\left[\sum_{j\in\{L,N\}}e^{-\frac{I_Er_j^{\alpha_j}}{M_sC_j}}\sum^{N_{j}-1}_{m=0}\left(\frac{I_Er_j^{\alpha_j}}{M_sC_j}\right)^m\frac{A_j}{\Gamma(m+1)}\right]\right]
\nonumber\\
&\overset{(h)}{=}\mathbb{E}_{r_j}\left[\sum_{j\in\{L,N\}}\sum^{N_{j}-1}_{m=0}\left(\frac{r_j^{\alpha_j}}{M_sC_j}\right)^m\frac{A_j}{\Gamma(m+1)}(-1)^m\mathcal{L}^{(m)}_{I_E}\left(\frac{r_j^{\alpha_j}}{M_sC_j}\right)\right].
\end{align}
step $(g)$ holds, since the serving BS at $x^*$ can be a LOS or NLOS BS, and step $(h)$ is due to the Laplace transform property $t^nf(t)\overset{\mathcal{L}}{\leftrightarrow} (-1)^n\frac{d^n}{d s^n}\mathcal{L}_{f(t)}(s)$.

In the following, we derive the analysis result of $\mathcal{L}_{I_E}\left(s\right)$.
\begin{align}
&\mathcal{L}_{I_E}\left(s\right)=\mathbb{E}\left(\textrm{e}^{-s{\sum_{z\in\Phi_E}G_b(\theta)L(x^*,z)g_{x^*z}}}\right)\overset{(i)}{=}\textrm{exp}\left(\int^{+\infty}_0\left(\textrm{e}^{-\frac{s}{x}}-1\right)\Lambda_E(0,dx)\right)
\nonumber\\
&\overset{(k)}{=}\textrm{exp}\left(-\int^{+\infty}_0\Lambda_E(0,x)\frac{s}{x^2}\textrm{e}^{-\frac{s}{x}}dx\right)\overset{(v)}{=}
\textrm{exp}\left(\underset{T}{\underbrace{-\int^{+\infty}_0\Lambda_E\left(0,\frac{1}{z}\right)s\textrm{e}^{-sz}dz}}\right),
\label{LaplaceTrasformIE}
\end{align}
step $(i)$ is obtained by using the probability generating functional (PGFL), step $(k)$ is obtained by using  integration by parts, and step $(v)$ is obtained by the variable replacing $z=\frac{1}{x}$.
Then, we concentrate on deriving the analysis result of $T$ in (\ref{LaplaceTrasformIE}).
Substituting $\Lambda_E(0,\frac{1}{z})$ in (\ref{LambdaE}) into $T$, $T$ can be rewritten as
\begin{align}
T=&-s\left(2\pi\lambda_E\sum_{j\in\{L,N\}}q_j\left(\sum_{V\in\{M_s,m_s\}}\textrm{Pr}_{G_b}(V)\frac{\left(VC_j\right)^{\frac{2}{\alpha_j}}}{\alpha_j}
\sum_{m=0}^{N_j-1}\frac{1}{m!}\underset{H_1}{\underbrace{\int^{\infty}_0\frac{\gamma\left(m+\frac{2}{\alpha_j},\frac{D^{\alpha_j}z}{VC_j}\right)}{z^{\frac{2}{\alpha_j}}}
\textrm{e}^{-sz}dz}}
\right)\right.
\nonumber\\
&\left.
\sum_{V\in\{M_s,m_s\}}\textrm{Pr}_{G_b}(V)\frac{\left(VC_N\right)^{\frac{2}{\alpha_N}}}{\alpha_N}
\sum_{m=0}^{N_N-1}\frac{1}{m!}\underset{H_2}{\underbrace{\int^{\infty}_0\frac{\Gamma\left(m+\frac{2}{\alpha_N},\frac{D^{\alpha_N}z}{VC_N}\right)}{z^{\frac{2}{\alpha_N}}}
\textrm{e}^{-sz}dz}}\label{T}
\right).
\end{align}

Finally, using \cite[eq. (6.455.1)]{Table} and \cite[eq. (6.455.2)]{Table}, the integral terms $H_1$ and $H_2$ can be calculated as
\begin{align}
H_1=\frac{\left(D^{\alpha_j}/(VC_j)\right)^{m+\frac{2}{\alpha_j}}\Gamma(m+1)}{\left(m+\frac{2}{\alpha_j}\right)\left(s+D^{\alpha_j}/(VC_j)\right)^{m+1}}
{_2}F_1\left(1,m+1;m+\frac{2}{\alpha_j}+1;\frac{D^{\alpha_j}/(VC_j)}{s+D^{\alpha_j}/(VC_j)}\right),
\label{H1}\\
H_2=\frac{\left(D^{\alpha_N}/(VC_N)\right)^{m+\frac{2}{\alpha_N}}\Gamma(m+1)}{\left(1-\frac{2}{\alpha_N}\right)\left(s+D^{\alpha_N}/(VC_j)\right)^{m+1}}
{_2}F_1\left(1,m+1;2-\frac{2}{\alpha_N}+1;\frac{S}{s+D^{\alpha_j}/(VC_j)}\right).\label{H2}
\end{align}
Finally, substituting (\ref{H1}) and  (\ref{H2}) into (\ref{T}), the closed-form result of $\mathcal{L}_{I_E}\left(s\right)$ can be obtained.
\section{Proof of Theorem 6}
Using total probability theorem, we have
\begin{align}
\textrm{Pr}\left(\textrm{SINR}_U\geq T_c\right)&= A_L\underset{Q_1}{\underbrace{\textrm{Pr}\left(\textrm{SINR}_U\geq T_c|\textrm{Serving BS is a LOS BS}\right)}}
\nonumber\\
&+A_N\underset{Q_2}{\underbrace{\textrm{Pr}\left(\textrm{SINR}_U\geq T_c|\textrm{Serving BS is a NLOS BS}\right)}}.
\end{align}

In the following, we detail the calculation of the conditional probability $Q_1$. The conditional probability $Q_2$ can be calculated with a similar procedure which is omitted for brevity. In the following, we denote the LOS terms in $\Phi_i$ as $\Phi_{i,L}$ and the NLOS terms in $\Phi_i$ as $\Phi_{i,N}$, for $i=I,A$.
\begin{align}
&Q_1=
\textrm{Pr}\left(h_{x^*o}\geq \frac{T_cr_L^{\alpha_L}\left({\sum_{y\in\Phi_I}M_s\phi P_th_{yx^*}L(y,x^*)+\sum_{y\in\Phi_A}M_a(1-\phi) P_th_{xo}L(y,x^*)}+N_0\right)}{\phi P_t M_sC_L}\right)\overset{(z)}\lessapprox
\nonumber\\
&1-\mathbb{E}\left(1-\textrm{exp}\left(-{\frac{a_LT_cr_L^{\alpha_L}}{\phi P_t M_sC_L}\left({\sum_{y\in\Phi_I}M_s\phi P_th_{yx^*}L(y,x^*)+\sum_{y\in\Phi_A}M_a(1-\phi) P_th_{yx^*}L(y,x^*)+N_0}\right)}\right)\right)^{N_L}\nonumber\\
%&=\mathbb{E}_{r_L}\left(\sum_{n=1}^{N_L}(-1)^{n+1}\textrm{exp}\left(-\Theta_L(n)r_L^{\alpha_L}N_0\right)
%\prod_{i\in\{L,N\}}\mathbb{E}_{\Phi_{I,i}}\left(\textrm{exp}\left(-\Theta_L(n)r_L^{\alpha_L}\sum_{y\in\Phi_{I,i}}M_s\phi P_th_{xo}L(y,x^*)\right)\right)\right.
%\nonumber\\
%&\left.\prod_{i\in\{L,N\}}\mathbb{E}_{\Phi_{A,i}}\left(\textrm{exp}\left(-\Theta_L(n)r_L^{\alpha_L}\sum_{y\in\Phi_{A,i}}M_a(1-\phi) P_th_{yx^*}L(y,x^*)\right)\right)\right).
%\nonumber\\
&\overset{(x)}{=}\mathbb{E}_{r_L}\left(\sum_{n=1}^{N_L}(-1)^{n+1}\binom{N_L}{n}\textrm{exp}\left(-\Theta_L(n)r_L^{\alpha_L}N_0\right)\prod_{j\in\{L,N\}}\mathbb{E}_{\Phi_{I,j}}\left(\prod_{y\in\Phi_{I,j}}\left(1+\Theta_L(n)r_L^{\alpha_L}M_s\phi P_tL(y,x^*)\right)^{-N_j}\right)\right.
\nonumber\\
&\left.\prod_{j\in\{L,N\}}\mathbb{E}_{\Phi_{A,j}}\left(\prod_{y\in\Phi_{A,j}}\left(1+\Theta_L(n)r_L^{\alpha_L}M_a(1-\phi) P_tL(y,x^*)\right)^{-N_j}\right)\right).\label{prooftheorem6LOS}
\end{align}
Step $(z)$ is due to the tight lower bound of the CDF of the gamma random variable given in \cite{Inequality}; step $(x)$ is due to the multinomial expansion and  Laplace transform of the gamma random variable.

On the condition that the serving BS is a LOS BS and the distance from the serving BS to the typical user is $r_L$,
from the user association policy in Section II-F, we know that
the nearest distance from the interfering BS in $\Phi_{I,L}$ and $\Phi_{A,L}$ to the typical authorized user should be larger than $r_L$, and the nearest distance from the interfering BS in $\Phi_{I,N}$ and $\Phi_{A,N}$ to the typical authorized user  should be larger than $\left(\frac{C_N}{C_L}\right)^{\frac{1}{\alpha_N}}r_L^{\frac{\alpha_L}{\alpha_N}}$.
Then, using PGFL, we have
\begin{align}
&\mathbb{E}_{\Phi_{I,L}}\left(\prod_{y\in\Phi_{I,L}}\left(1+\Theta_L(n)r_L^{\alpha_L}M_s\phi P_tL(y,x^*)\right)^{-N_L}\right)=\varpi_1(M_s,\phi,n)
\nonumber\\
&\mathbb{E}_{\Phi_{I,N}}\left(\prod_{y\in\Phi_{I,N}}\left(1+\Theta_L(n)r_L^{\alpha_L}M_s\phi P_tL(y,x^*)\right)^{-N_N}\right)=\varpi_2(M_s,\phi,n)\varpi_3(M_s,\phi,n)
\end{align}
Similarly, the analysis resut of $\mathbb{E}_{\Phi_{A,j}}\left(\prod_{y\in\Phi_{A,j}}\left(1+\Theta_L(n)r_L^{\alpha_L}M_a(1-\phi) P_tL(y,x^*)\right)^{-N_j}\right)$, $j\in\{L,N\}$ can be obtained.
Finally, substituting the pdf of $r_L$ into (\ref{prooftheorem6LOS}), the proof can be completed.

\section{Proof of Theorem 7}
In the following, for the convenience of expression, we denote the set of the LOS eavesdroppers in $\Phi_Z$ as $\Phi_{Z,L}$, and denote the set of the NLOS eavesdroppers as $\Phi_{Z,N}$.

Since the AN signals received at multiple eavesdroppers are not independent, we resort to the technique in \cite{Enhancing} to derive a lower bound of $p_{sec}$. We first derive the conditional secrecy probability conditioned on $\Phi_A$. Then, with the Jensen's inequality, we derive a lower bound of $p_{sec}$. Specially
\begin{align}
p_{sec}
=&\mathbb{E}_{\Phi_{A}}\left(\mathbb{E}_{\Phi_{Z,L},\Phi_{Z,N}}\left(\left.\prod_{z\in\Phi_{Z,L}}\mathrm{Pr}\left(\textrm{SINR}_z\leq \mathrm{T}_e|\Phi_{A}\right)\prod_{z\in\Phi_{Z,N}}\mathrm{Pr}\left(\textrm{SINR}_z\leq \mathrm{T}_e|\Phi_{A}\right)\right.\right)
\right)
\nonumber\\
=&\mathbb{E}_{\Phi_{A}}\left(\textrm{exp}\left(-C\lambda_E\textrm{Pr}_{x}(M_s)\int_{\mathbb{R}^2\cap B(o,D)}\mathrm{Pr}\left(\textrm{SINR}_z\geq \textrm{T}_e|z\in\Phi_{Z,L},\Phi_{A}\right)dz\right.\right.
\nonumber\\
&\qquad\qquad\quad-(1-C)\lambda_E\textrm{Pr}_{x}(M_s)\int_{\mathbb{R}^2\cap B(o,D)}\mathrm{Pr}\left(\textrm{SINR}_z\geq \textrm{T}_e|z\in\Phi_{Z,N},\Phi_{A}\right)dz
\nonumber
\\
&\left.\left.\qquad\qquad\quad-\lambda_E\textrm{Pr}_{x}(M_s)\int_{\mathbb{R}^2/ B(o,D)}\mathrm{Pr}\left(\textrm{SINR}_z\geq \textrm{T}_e|z\in\Phi_{Z,N},\Phi_{A}\right)dz\right)\right)\nonumber
\\
\overset{(n)}{\gtrapprox} & \textrm{exp}\left(-2\pi C\lambda_E\textrm{Pr}_{x}(M_s)\int^D_{0}\mathrm{Pr}\left(\textrm{SINR}_z\geq \textrm{T}_e|z\in\Phi_{Z,L}\right)rdr\right)
\nonumber\\
&\textrm{exp}\left(-2\pi(1-C)\lambda_E\textrm{Pr}_{x}(M_s)\int^D_{0}\mathrm{Pr}\left(\textrm{SINR}_z\geq \textrm{T}_e|z\in\Phi_{Z,N}\right)rdr\right)
\nonumber\\
&\textrm{exp}\left(-2\pi\lambda_E\textrm{Pr}_{x}(M_s)\int^{+\infty}_{D}\mathrm{Pr}\left(\textrm{SINR}_z\geq \textrm{T}_e|z\in\Phi_{Z,N}\right)rdr\right),\label{ProofTheorem7processpsec}
\end{align}
where $r$ denotes the link length between the eavesdropper and the serving BS. We should point out that the conditional probabilities in the first line of the equation (\ref{ProofTheorem7processpsec}) denote the probabilities
that the SINR received by eavesdroppers at different positions is not larger than T$_e$, conditioned on a common $\Phi_{A}$.
Changing to a polar
coordinate system and using Jensen's inequality, we can obtain step $(n)$.

In the following, we first show the derivation of $\textrm{Pr}\left(\textrm{SINR}_z\geq \textrm{T}_e|z\in\Phi_{Z,L}\right)$. Then,
similar procedures can be adopted to derive $\textrm{Pr}\left(\textrm{SINR}_z\geq \textrm{T}_e|z\in\Phi_{Z,N}\right)$, which are omitted, for brevity.
\begin{align}
&\textrm{Pr}\left(\textrm{SINR}_z\geq \textrm{T}_e|z\in\Phi_{Z,L}\right)=
\textrm{Pr}\left(g_{x^*z}\geq \frac{r^{\alpha_L}\textrm{T}_e\left(I_{A_z}+N_0\right)}{\phi P_tM_sC_L}|z\in\Phi_{Z,L}\right)\overset{(v)}{\lessapprox}
\nonumber\\
& 1-\mathbb{E}_{\Phi_{A}}\left(1-\textrm{e}^{-\frac{\textrm{T}_er^{\alpha_L}a_L\left(I_{A_z}+N_0\right)}{\phi P_tM_sC_L}}\right)^{N_L}
=\sum^{N_i}_{n=1}(-1)^{n+1}\binom{N_L}{n}\textrm{e}^{-\frac{na_L\textrm{T}_eN_0r^{\alpha_L}}{\phi P_tM_sC_L}}
\mathcal{L}_{I_{A_z}}\left({\frac{na_L\textrm{T}_er^{\alpha_L}}{\phi P_tM_sC_L}}\right),\label{ProofTheorem7process}
%\mathbb{E}_{\Phi_{A}}\left(\textrm{e}^{-\frac{na_L\textrm{T}_er^{\alpha_L}I_{A_z}}{\phi P_tM_sC_L}}\right),\label{ProofTheorem7process}
\end{align}
Since $g_{x^*z}\sim\textrm{gamma}\left(N_L,1\right)$, with the inequality in \cite{Inequality},  step $(v)$ can be obtained.

For facilitating the derivations, we first introduce the following definition.
\begin{definition}
The path loss process with fading (PLPF) $\mathcal{N}_z$ is the point process on $\mathbb{R}^+$ mapped from $\Phi_{A}$, where $\mathcal{N}_z\triangleq\left\{\zeta_y=\frac{1}{M_ag_{yz}L(y,z)},y\in\Phi_A\right\}$ and $z\in\mathbb{R}^2$. We sort the elements of $\mathcal{N}_z$ in ascending order and introduce the index such that $\zeta_i\leq \zeta_j$ for $\forall i<j$.
\end{definition}

Then, following the proof of Lemma 4, the intensity measure of $\mathcal{N}_z$ can be calculated as
\begin{align}
\Lambda_z\left(0,t\right)=&2\pi\lambda_B\sum_{j\in\left\{L,N\right\}}q_j\textrm{Pr}_{x}(M_a)\frac{\left(M_aC_jt\right)^{\frac{2}{\alpha_j}}}{\alpha_j}\sum^{N_j-1}_{m=0}\frac{\gamma\left(m+\frac{2}{\alpha_j},\frac{D^{\alpha_j}}{M_aC_jt}\right)}{m!}
\nonumber\\
&+2\pi\lambda_B\textrm{Pr}_{x}(M_a)\frac{\left(M_aC_Nt\right)^{\frac{2}{\alpha_N}}}{\alpha_N}\sum^{N_N-1}_{m=0}\frac{\Gamma\left(m+\frac{2}{\alpha_N},\frac{D^{\alpha_N}}{M_aC_Nt}\right)}{m!},
\nonumber
\end{align}
where $q_L=C$ and $q_N=1-C$.

The the Laplace transform $\mathcal{L}_{I_{A_z}}\left(s\right)$ can be calculated as
%=\mathbb{E}\left(\textrm{exp}\left(-s\left(1-\phi\right)P_t{\sum_{y\in\Phi_{A_z}}M_ag_{y,z}L(y,z)}\right)\right)
\begin{align}
&\mathcal{L}_{I_{A_z}}\left(s\right){=}\textrm{exp}\left(\int^{+\infty}_0\left(\textrm{e}^{-\frac{s\left(1-\phi\right)P_t}{x}}-1\right)\Lambda_z(0,dx)\right)
{=}\textrm{exp}\left(-\int^{+\infty}_0\Lambda_z(0,x)\frac{s\left(1-\phi\right)P_t}{x^2}\textrm{e}^{-\frac{s\left(1-\phi\right)P_t}{x}}dx\right)\nonumber\\
&\overset{(k)}{=}
\textrm{exp}\left({{-\int^{+\infty}_0\Lambda_z\left(0,\frac{1}{z}\right)s\left(1-\phi\right)P_t\textrm{e}^{-s\left(1-\phi\right)P_tz}dz}}\right)
\end{align}
Step $(k)$ is obtained by the variable replacing $z=\frac{1}{x}$. Following the derivation of the analysis result of (\ref{LaplaceTrasformIE}) in the proof of Theorem 3, we can derive the analysis result of $\mathcal{L}_{I_{A_z}}\left(s\right)$.
Substituting  the analysis result of $\mathcal{L}_{I_{A_z}}\left(s\right)$ into (\ref{ProofTheorem7process}), the analysis result of $\textrm{Pr}\left(\textrm{SINR}_z\geq \textrm{T}_e|z\in\Phi_{Z,L}\right)$ can be obtained.

%Finally, substituting the analysis result of $\mathbb{E}_{\Phi_{A_z}}\left(\textrm{Pr}\left(\textrm{SINR}_z\geq \textrm{T}_e|z\in\Phi_{Z,L}\right)\right)$ into (\ref{ProofTheorem7processpsec}), the proof can be completed.

\end{document}